\documentclass[10pt]{article}
\usepackage{amssymb}
\usepackage{amscd}
\usepackage{latexsym}
\usepackage{endnotes}
\usepackage{graphicx}
\renewcommand{\theequation}{\arabic{section}.\arabic{equation}}
\newcommand{\newsec}{\setcounter{equation}{0}\section}%
\newcommand{\Zz}{{\mathbb Z}}

\newcommand{\Rr}{{\mathbb R}}
\newcommand{\Cc}{{\mathbb C}}
\newcommand{\Tt}{{\mathbb T}}

\def\be{\begin{equation}}
\def\ee{\end{equation}}
\def\bea{\begin{eqnarray}}
\def\eea{\end{eqnarray}}

\def\d{{\,\rm d}}

\def\0{{\bf 0}}
\def\F{{\bf F}}
\def\a{{\bf a}}

\def\k{{\bf k}}
\def\e{{\bf e}}

\def\q{{\bf q}}
\def\l{{\bf l}}
\def\m{{\bf m}}
\def\n{{\bf n}}
\def\r{{\bf r}}
\def\p{{\bf p}}

\def\v{{\bf v}}
\def\x{{\bf x}}
\def\y{{\bf y}}
\def\z{{\bf z}}
\def\t{{\bf t}}
\def\K{{\bf K}}

\def\P{{\bf P}}

\def\veps{\varepsilon}
\def\h2m{\frac{\hbar^2}{2m}}
\def\p0{{P_{\beta H^0_N}}}

\def\calm{{{\cal M}_\Lambda}}
\def\calmst{{\calm^{\rm st}}}
\def\supp{{\rm supp\,}}

\newtheorem{theorem}{Theorem}[section]

\newtheorem{lemma}{Lemma}[section]
\newtheorem{definition}{Definition}[section]
\newtheorem{proposition}{Proposition}[section]
\newtheorem{corollary}{Corollary}[section]
\newtheorem{conjecture}{Conjecture}[section]

\begin{document}
\title{{\flushleft{\small {\rm Published in Commun. Math. Phys. \textbf{305}, 657-710 (2011)}\\}}\vspace{1cm}
\large\bf Ground state at high density}
\author{Andr\'as S\"ut\H o\\Research Institute for Solid State Physics and Optics\\Hungarian Academy of Sciences\\P. O. B. 49, H-1525 Budapest, Hungary\\
E-mail: suto@szfki.hu\\}
\date{}
\maketitle
\thispagestyle{empty}
\begin{abstract}
\noindent
Weak limits as the density tends to infinity of classical ground states of integrable pair potentials are shown to minimize the mean-field energy functional. By studying the latter we derive global properties of high-density ground state configurations in bounded domains and in infinite space. Our main result is a theorem stating that for interactions having a strictly positive Fourier transform the distribution of particles tends to be uniform as the density increases, while high-density ground states show some pattern if the Fourier transform is partially negative. The latter confirms the conclusion of  earlier studies by Vlasov (1945), Kirzhnits and Nepomnyashchii (1971), and Likos \emph{et al.} (2007). Other results include the proof that there is no Bravais lattice among high-density ground states of interactions whose Fourier transform has a negative part and the potential diverges or has a cusp at zero. We also show that in the ground state configurations of the penetrable sphere model particles are superimposed on the sites of a close-packed lattice.

\vspace{2mm}
\noindent
PACS: 61.50.Ah, 02.30.Nw, 61.50.Lt

\vspace{2mm}
\noindent\textbf{Contents}

\renewcommand\theenumii{\arabic{enumi}.\arabic{enumii}}
\renewcommand\labelenumii{\theenumii}

\begin{enumerate}
\item
Introduction
\item
Best superstability constant
\item
Infinite-density limit of the ground state energy and the free energy per pair
\item
Infinite-density ground state and the energy functional in finite volume
\begin{enumerate}
\item
Bounded interactions
\item
Unbounded interactions
\item
Fourier representation of the energy functional
\end{enumerate}
\item
Stability conditions for pair potentials
\item
Stationary points of the energy functional
\item
Ground state configurations in infinite space
\item
Infinite-density ground state in infinite space
\item
Infinite-density ground states for $v\geq 0$
\item
Uniformity vs nonuniformity of high-density ground states
\item
Interactions without Bravais lattice ground states at high density
\begin{enumerate}
\item
Compensation by higher harmonics
\item
Potentials with a cusp at zero
\end{enumerate}
\item
The penetrable sphere model
\end{enumerate}
\end{abstract}

\newsec{Introduction}

The present paper is a continuation of my earlier work on ground states of bounded Fourier-transformable pair potentials~\cite{Su1,Su2}. Bounded or integrable interactions appear in quantum physics for example in Bogoliubov's theory of the Bose gas, which is based on models expressed in terms of the Fourier transform of the pair potential. Such interactions play also a central role in classical soft matter physics~\cite{Li1}. Those studied in~\cite{Su1,Su2} had a nonnegative Fourier transform of compact support. In 2007 Likos and his coworkers published a mean-field study of the case when the Fourier transform of the pair potential has a negative part~\cite{Lik}. If the system discussed in \cite{Su1,Su2} showed already very peculiar properties at high densities, according to these authors a partly negative Fourier transform induced an even more curious behavior, the particles having the tendency to form clusters on the sites of a lattice whose lattice constant would be determined by the (negative) minimum of the Fourier transform, and only the population of the clusters would increase with the density. The question then arises whether it is possible to prove rigorously this kind of behavior. When, in the fall of 2008, I presented preliminary results on this problem and on its quantum mechanical counterpart \cite{Su3} at a Montreal meeting in mathematical physics, Valentin Zagrebnov kindly informed me that, without knowing it, I worked on the theory of coherent crystals, a subject promoted by Russian physicists a long time ago.

In effect, maybe the first attempt at a theory of crystallization was due to Vlasov. In his paper~\cite{Vla} Vlasov studied the solutions of the today called Vlasov equation
\be\label{Vlasov}
-\frac{\partial f}{\partial t}=\v\cdot\frac{\partial f}{\partial\r}+\frac{1}{m}\F\cdot\frac{\partial f}{\partial\v}
\ee
for the one-particle distribution $f(\r,\v,t)$. There is no collision term, but the equation contains a self-consistent force field

\be
\F(\r,t)=-\frac{\partial}{\partial\r}\int u(\r-\r')\int f(\r',\v,t)\d\v\d\r'
\ee
induced by a translation invariant pair potential $u$. Considered on the torus or in the entire space, this equation has spatially homogeneous static solutions $f^0(\v)$. Vlasov investigated his equation after linearizing it about $f^0$. Besides many others, he asked the question whether there may exist spatially periodic static solutions. If yes, they could be associated with crystals. In his analysis, however, he arrived at the false conclusion that for the existence of such a solution $\int u(\r)\d\r<0$ must hold. Today we know that there can be no equilibrium (crystalline or other) phase under the effect of such an interaction: the system collapses, the grand-canonical partition function diverges in finite volumes~\cite{Rue}. The possibility of a static periodic solution for a (super)stable $u$ is, nonetheless, there, and to see how such a solution emerges, we can follow Vlasov's reasoning almost up to the end. Given the velocity profile $f^0$, one substitutes $f(\r,\v,t)=f^0(\v)+\phi(\r,\v,t)$ into Eq.~(\ref{Vlasov}) and keeps only the terms linear in $\phi$:
\be
\frac{\partial\phi}{\partial t}+\v\cdot\frac{\partial\phi}{\partial \r}-\frac{1}{m}\frac{\partial f^0}{\partial \v}\cdot\frac{\partial (u*\Phi)}{\partial \r}=0,
\ee
where
\be
\Phi(\r,t)=\int\phi(\r,\v,t)\d\v.
\ee
With the ansatz
\be
\phi(\r,\v,t)=e^{i\omega t-i\k\cdot\r}g_\k(\v),
\ee
one obtains
\be
g_\k(\v)=\frac{\widehat{u}(\k)\k\cdot\frac{\partial f^0}{\partial\v}}{m(\k\cdot\v-\omega)}\int g_\k(\v')\d\v',
\ee
where $\widehat{u}$ is the Fourier transform of $u$. Integration over $\v$ yields
\be
1=\frac{\widehat{u}(\k)}{m}\int\frac{\k\cdot\frac{\partial f^0}{\partial\v}}{\k\cdot\v-\omega}\d\v
\ee
which implicitly determines the dispersion relation $\omega(\k)$. With the choice $f^0(\v)=h(\v^2)$ this further simplifies, and for a static solution ($\omega=0$), $\k$ must satisfy the equation
\be\label{cond-k-0}
1=\frac{2\widehat{u}(\k)}{m}\int h'(\v^2)\d\v.
\ee
In three dimensions, for a bounded and sufficiently fast decaying $h$ this becomes
\be\label{cond-k}
1=-2\pi\frac{\widehat{u}(\k)}{m}\int_0^\infty\frac{h(x)}{\sqrt{x}}\d x.
\ee
Because $h\geq 0$, we see that a solution for $\k$ is possible only if $\widehat{u}$ has a negative part! Making the substitutions $f^0(\v)=h(\v^2)$ and $\omega=0$ also in $g_\k$, we obtain an approximate static solution of Eq.~(\ref{Vlasov}) in the form
\be\label{Vla-per}
f(\r,\v)=h(\v^2)+h'(\v^2)\sum c_\k \cos\k\cdot\r,
\ee
with the sum running over the vectors $\k$ which solve Eq.~(\ref{cond-k}), and the real $c_\k$ (incorporating the constant $(2/m)\widehat{u}(\k)\int g_\k$) chosen so that $f\geq 0$. To obtain (\ref{Vla-per}) we have supposed that $u(\x)=u(-\x)$, which implies $\widehat{u}(\k)=\widehat{u}(-\k)$ real and (\ref{cond-k}) holding simultaneously for $\pm\k$. If $u$ is integrable, then $\widehat{u}(\k)$ is continuous and decays at infinity; thus, for any ``natural'' $u$ the sum in (\ref{Vla-per}) is finite. Moreover, $\widehat{u}(\0)>0$ for a superstable interaction, therefore $\widehat{u}(\k)\leq 0$ can be only if $|\k|\geq k_0>0$. Let now $f^0$ be the Maxwell distribution, i.e.,
\be
h(x)=\rho\left(\frac{\beta m}{2\pi}\right)^{3/2}e^{-\beta mx/2},
\ee
where $\rho$ is the density and $\beta$ is the inverse temperature. Then $h'(x)=-(\beta m/2)h(x)$, and (\ref{cond-k-0}) becomes
\be
-\beta\rho\widehat{u}(\k)=1.
\ee
This equation has a solution if $\beta\rho\geq (\beta\rho)_c$, where
\be
(\beta\rho)_c=-\frac{1}{\min\{\widehat{u}(\k)\}}.
\ee
If the minimum of $\widehat{u}$ is nondegenerate, at the critical point the resulting distribution is spatially periodic. As $\beta\rho$ increases, the solutions for $\k$ shift away from the minimizer of $\widehat{u}$, and the solution of the linearized equation is, in general, almost periodic. With the Maxwell distribution (\ref{Vla-per}) becomes
\be
f(\r,\v)=f^0(\v)\left[1-\frac{\beta m}{2}\sum c_\k\cos\k\cdot\r\right].
\ee
In order to keep $f$ nonnegative, as $\beta$ increases, the coefficients $c_\k$ must scale as $\beta^{-1}$.

The right condition for the formation of ``coherent crystals'' appeared only much later, in papers by Kirzhnits and Nepomnyashchii~\cite{KN,N}. The term referred to (hypothetical) crystals of mobile particles as in Vlasov's theory, capable of ballistic motion which must be coherent if it preserves long-range order. The context was somewhat different, these authors were interested in the crystallization of a quantum liquid. The starting point was a Hamiltonian the ground state of which was determined in the Hartree approximation. This analysis showed that the lowest-energy solution of the Hartree equation can be periodic only if the Fourier transform of the pair potential is partially negative, and the periodicity is then given by the wave vector at which the Fourier transform is minimal. These authors, just as Vlasov, were fully aware of the extraordinary properties of coherent crystals, even more pronounced in the quantum than in the classical case, and described them almost in the same terms as Likos \emph{et al.}~\cite{Lik}.

The mean-field or density-functional approach of Likos \emph{et al.}, the partial differential equation method of Vlasov, and the Hartree approximation of Kirzhnits and Nepomnyashchii are effective one-particle theories, and all arrive at the same conclusion. One can have little doubt in their truth. The problem is more difficult than the case treated in \cite{Su1,Su2}, and a new idea is necessary to obtain some progress in the rigorous theory. The idea presented and exploited in this paper is that of an infinite-density ground state (IDGS). Working in a finite domain (on a torus here), the $N$-particle ground states are arrangements of $N$ points that minimize the interaction energy. To each $N$-point subset one can assign a discrete measure, and an IDGS is the weak limit of a sequence of discrete measures associated with $N$-particle ground states, when $N$ goes to infinity. The ground state energy and energy per particle diverge in this limit but, if the interaction is integrable, the energy per pair is convergent. The limit is proportional to the best superstability constant, the largest number that can multiply $N^2/V$ in a lower bound on the energy of $N$ particles in a cube of volume $V$. The two notions, that of an IDGS and of the best superstability constant, can be related via an energy functional written for normalized measures on the torus. It is shown that any IDGS is a minimizer of the energy functional, and the value of the minimum is the best superstability constant. The task is then to find these minimizers, because they can provide information on ground state configurations at high but finite densities. Most results on IDGS's will be obtained by writing the energy functional in Fourier representation; an exception is the penetrable sphere model presented at the end of the paper (for another example see Ref.~\cite{Su4}). Our main result is a confirmation of the conclusion of the earlier works~\cite{Lik,Vla,KN,N}: if the Fourier transform of the pair potential has a negative part, the distribution of particles in high-density ground states is nonuniform. In some cases, e.g. when the potential diverges at the origin or has a cusp there, this nonuniform distribution cannot be approached through Bravais lattices. Except for the penetrable sphere model, from this analysis we cannot predict the ground state configurations at high densities with the precision obtained in Refs.~\cite{Su1,Su2}. The lack of precise information is true also for interactions with a strictly positive Fourier transform, but we can at least assert that, in contrast with the former, the asymptotic distribution of particles as the density increases tends to be uniform. Since there is a continued interest in the Gaussian core model~\cite{Stil}-\cite{CKS}, even this weak result may be of some value.

It is to be emphasized that the limit of infinite \emph{particle} density is not at all unusual in classical physics. Whenever a continuum theory is applied to describe a system of classical pointlike particles, tacitly this limit is used. Theories of classical fluids, the rigorous van der Waals theory of liquid-vapor phase transition~\cite{LP,GP,BBP}, the continuum theories of droplet formation~\cite{CCE1,CCE2} and liquid-gas interfaces~\cite{Mod} are based on different free energy functionals defined on continuous mass densities. These theories can be derived through scaling limits in which the particle density tends to infinity while the mass of the particles and the interactions among them tend to zero in order to keep the mass and energy densities finite. In this work we do not have to scale the mass because it enters only the kinetic energy which vanishes in classical ground states (gravitating systems are out of the scope of this study), but we do scale the interaction: considering the energy per pair of particles corresponds to scaling the pair potential by dividing it with $N$ as $N$ tends to infinity in a fixed volume. The crucial difference compared with the continuum theories cited above is the extension of the energy functional to discrete distributions. Starting with them and by controlling their convergence in the limit of infinite density we can obtain information on ground state configurations at finite densities.

The paper is organized as follows. Section~\ref{best} fixes the conditions on the interaction and the basic notations, and contains the definition of the best superstability constant together with some preliminary results on it. Throughout the paper we deal with both bounded and unbounded interactions. The results will be more complete for bounded interactions, and the reason of this appears already in Sections~\ref{best} and \ref{idens-limit}. For bounded interactions the self-energy is finite, and the energy per pair with and without the self-energy converges to the best superstability constant as $N$ goes to infinity from above and from below, respectively. This permits to prove that the convergence is uniform in the volume -- a fact that we do not know for unbounded interactions. In Section~\ref{idens-limit} we still show that the infinite-density limit of the free energy per pair is independent of the temperature and is the same as that of the ground state energy. In Section~\ref{sec-IDGS-dens-func} we define the infinite-density ground states as weak limits of Dirac combs associated with $N$-particle ground states, introduce the energy functional and prove two theorems, one for bounded and another for unbounded interactions. They show that IDGS's minimize the energy functional, and the minimum is the best superstability constant. From the Fourier representation of the energy functional, introduced in Section~\ref{sec-Fourier}, we derive some stability conditions in Section~\ref{sec-stability}. Section~\ref{sec-stat} contains a complete description of the stationary points of the energy functional. Their knowledge is important because there can be minimizers among them. The truly challenging problem is to find the ground state configurations (GSC) in infinite space. The definition, based on local stability, is recalled in Section~\ref{sec-GSC}. Here we prove two lemmas, the first relating periodic configurations and minimizers of the energy density, the second establishing the relation between periodic GSC's in infinite space and GCS's on tori. The analysis of the problem in infinite space is continued in Section~\ref{sec-IDGS-infinite}. We introduce the notion of an IDGS in infinite space and formulate as a conjecture a commutative diagram relating four objects: GSC and IDGS in finite domains and in infinite space. The main result of this section is Proposition~\ref{prop-C[u]} which gives a new expression of the best superstability constant. Sections~\ref{subsec-vgeq0} -- \ref{sec-penetrable} present applications. In Section~\ref{subsec-vgeq0} we give a rather complete description of IDGS's in the case when the Fourier transform of the interaction is nonnegative. Section~\ref{sec-unif-nonunif} contains the main theorem proving the asymptotic uniformity or non-uniformity of GSC's in infinite space for the case of a strictly positive or partially negative Fourier transform, respectively. In Section~\ref{sec-noBravais} we discuss two special classes of interactions with a partly negative Fourier transform. In the first one for any nonzero $\k$ the sum of the Fourier transform over integer multiples of $\k$ is nonnegative. In the second case the Fourier transform is ultimately positive and slowly decaying, so that the pair potential either diverges at zero or has a cusp there. In both cases we show that no Bravais lattice can be an IDGS, and high-density GSC's are different from Bravais lattices. On the contrary, bounded pair potentials that are flat or nesting at the origin or have a dominantly negative Fourier transform at large wave vectors may prefer the accumulation of particles on the sites of a lattice which is independent of the density. In Section~\ref{sec-penetrable} we demonstrate this property on the so-called penetrable sphere model in which the interaction is a repulsive square core potential. The paper ends with an Appendix.

\newsec{Best superstability constant}\label{best}

Consider the problem of the ground state of a system of classical identical particles confined in a fixed bounded domain $\Lambda$ of volume $V$ and interacting via a translation-invariant pair interaction $u(\x-\y)=u(\y-\x)$. Assume $u$ to be integrable, bounded outside the origin, strongly tempered, superstable and lower semicontinuous (see below). In the simplest case $\Lambda$ is a $d$-dimensional cube of side length $L$ taken with periodic boundary conditions, and $u$ is also made periodic. The usual way to achieve this is to replace $u(\x)$ by
\be\label{uper}
u_\Lambda(\x)=\sum_{\n\in\Zz^d}u(\x+L\n).
\ee
This series is absolutely convergent for any $L>0$ and any $\x$ (outside the set $L\Zz^d$ when $u(\0)=\infty$) if $u$ is strongly tempered. The general definition of strong temperedness is
\be\label{str}
|u(\x)|<C|\x|^{-d-\eta}
\ee
for $|\x|>r_0$, with some $r_0,C,\eta>0$. If $u$ has no hard core, $r_0=0$ can be taken. Condition (\ref{str}) also guearantees that $u_\Lambda$ converges to $u$ pointwise as $L$ tends to infinity. A ground state of $N$ particles in $\Lambda$ is any $N$-point configuration $(\r)_N=(\r_1,\ldots,\r_N)$ minimizing the $N$-particle interaction energy
\be\label{Ulambda}
U_\Lambda(\r)_N=\sum_{i<j}u_\Lambda(\r_i-\r_j).
\ee
The minimum energy will be denoted by $E_0(N)$. Because of the shift-invariance of $u$, any translate of a ground state is a ground state. To make sure that (\ref{Ulambda}) can be minimized for all $N$, we will ask $u$ to be lower semicontinuous~\cite{Rud}.

A way to understand the behavior of soft matter at high density is to study the limit of infinite density. Since the particles have no hard core, this limit can be given a meaningful definition. The limit $\rho=N/V\to\infty$ will be realized by letting $N$ diverge in a fixed volume. To characterize a ground state, a suitable scaling of $U_\Lambda$ is necessary. If $u$ is bounded then
\be\label{omega'rn}
\epsilon(\r)_N=\frac{V}{N(N-1)}U_\Lambda(\r)_N
\ee
remains bounded in this limit whatever the sequence of $N$-point configurations is. Moreover,
\be\label{omega'n}
\epsilon_N=\min_{(\r)_N\subset\Lambda}\epsilon(\r)_N= \frac{V}{N(N-1)}E_0(N)
\ee
is bounded as $N$ goes to infinity even if $u$ is unbounded but is integrable and bounded from below. Indeed, divide $\Lambda$ into $N$ cubes of volume $V/N$ and choose $(\r)_N$ to be the centers of the cubes. Then the second sum in
\be
U_\Lambda(\r)_N=\frac{N}{2V}\sum_{i}\sum_{j\neq i}u_\Lambda(\r_i-\r_j)\frac{V}{N}
\ee
is a Riemann-sum, therefore
\be\label{Rie}
\epsilon(\r)_N=\frac{1}{2}\int u(\x)\d\x+o(1)\quad(N\to\infty)
\ee
is an upper bound to $\epsilon_N$. We conclude that the right quantity to look at is $\epsilon(\r)_N$ that, with a slight abuse, we call the pair energy.

The search for the ground state in the limit of $N$ going to infinity is closely related to finding the \emph{best superstability constant} for $u$. The notion of superstability was introduced by Ruelle \cite{Rue1,Rue}. Stability means the existence of a constant $B$ such that for any $N$ the ground state energy is bounded below by $-BN$; superstability means that the ground state energy density (energy per volume) increases with the particle density at least quadratically. If $u$ is integrable and superstable, to leading order in the density the increase is quadratic.

\begin{definition} The best superstability constant for a superstable interaction $u$ is the supremum of the positive numbers $C$ such that for any large enough cube $\Lambda$ of volume $V$
\be\label{ss}
U_\Lambda(\r)_N\geq CN^2/V
\ee
holds for every $N$ above a possibly $\Lambda$-dependent value and every $(\r)_N\subset\Lambda$.
\end{definition}
\begin{lemma}\label{lem-Cu} The best superstability constant for $u$ is
\be\label{Cu}
C[u]=\liminf_{V\to\infty}C_\Lambda[u]
\ee
where the limit is taken over cubes of increasing volume $V$ and
\be\label{Clamu}
C_\Lambda[u]=\liminf_{N\to\infty}\epsilon_N.
\ee
\end{lemma}

\emph{Proof.} Take any $C<C[u]$. From the definition of liminf it follows that $C_\Lambda[u]>C$ for $\Lambda$ large enough, say, $\Lambda\supset\Lambda_C$. It also follows that there exists some $N_\Lambda$ such that
\be
\frac{N-1}{N}\,\epsilon_N>C\quad{\rm if}\quad\Lambda\supset\Lambda_C\quad{\rm and}\quad  N>N_\Lambda
\ee
which is (\ref{ss}). On the other hand, if $C>C[u]$ then there exist arbitrarily large domains $\Lambda$ such that $C>C_\Lambda[u]$, and an infinite sequence $N_j$ of positive integers (which may depend on $\Lambda$) such that
\be\label{conv}
\lim_{j\to\infty}\epsilon_{N_j}=C_\Lambda[u].
\ee
Thus,
\be\label{conv2}
\epsilon_{N_j}< C
\ee
for $j$ large enough, which is the negation of (\ref{ss}). $\square$

\noindent
\emph{Remarks.} (i) In the definition of superstability \cite{Rue} there is an additional term $-BN$ on the right-hand side of (\ref{ss}). We can add such a term with any $B\geq 0$ without violating the inequality, it even allows for the extension of the inequality to every $N$, cf. Eq.~(\ref{super}) below. Obviously, no choice of $B$ permits to increase the coefficient of $N^2$ and attain a superstability constant larger than $C[u]$ defined by (\ref{Cu}). (ii) Later on, $\Lambda$ will be a general parallelepiped. The periodized interaction can be defined, and the analogue of Lemma~\ref{lem-Cu} can be proven in the following form. Let $\Lambda_0$ be any nondegenerate parallelepiped, and for $s>0$ consider $s\Lambda_0=\{s\x:\x\in\Lambda_0\}$. Then the best superstability constant is $\liminf_{s\to\infty}C_{s\Lambda_0}[u]$. The limit actually exists,
\be\label{Cu-Fisher}
C[u]=\lim_{\Lambda\to\infty}C_{\Lambda}[u]
\ee
if $\Lambda$ tends to infinity in the Fisher sense~\cite{Rue}, and is the same as that one obtains with the use of $u$ instead of the periodized $u_\Lambda$. Based on the strong temperedness of the interaction, a proof in analogy with the proof of existence of the thermodynamic limit of the free energy~\cite{Rue} could be done, but in this paper (\ref{Cu-Fisher}) is considered as a hypothesis.

An integrable interaction $u$ has a bounded continuous Fourier transform decaying at infinity \cite{Rud}. We will denote it by $v$. Thus,
\be
v(\k)=\int u(\r)e^{-i\k\cdot\r}\d\r.
\ee
If $\sum_{\k\in\Lambda^*}|v(\k)|<\infty$ then $\sum_{\k\in\Lambda^*}v(\k)e^{i\k\cdot\r}$ is a continuous function and
\be\label{exp}
u_\Lambda(\r)=\frac{1}{V}\sum_{\k\in\Lambda^*}v(\k)e^{i\k\cdot\r}
\ee
almost everywhere; if $u$ is continuous, equality holds everywhere. Here $\Lambda^*=(2\pi/L)\Zz^d$ if $\Lambda$ is a cube of side length $L$. Superstability implies that $v(\0)>0$ and also $u(\0)>0$ and $u_\Lambda(\0)>0$.

\begin{proposition} $C_\Lambda[u]\leq v(\0)/2$ and thus $C[u]\leq v(\0)/2$.
\end{proposition}
\emph{Proof.} This follows from Eq.~(\ref{Rie}). An alternative proof is obtained by computing the average of the potential energy,
\be
\frac{1}{V^N}\int_{\Lambda^N}U_\Lambda(\r)_N\d(\r)_N=\frac{v(\0)N(N-1)}{2V}.
\ee
The minimum is smaller than the average,
\be\label{min-av}
\frac{N(N-1)}{V}\epsilon_N=E_0(N)\leq\frac{v(\0)N(N-1)}{2V}
\ee
and hence
\be\label{Cv0}
C_\Lambda[u]\leq v(\0)/2.\quad\Box
\ee

The following simple observation helps to see the possibility of superposition of particles in some ground state configurations. Given $(\r)_M$, for any integer $n\geq 1$ we define the $nM$-point configuration $(\r)_{M}^n$ by $n$ times repeating $(\r)_M$,
\be\label{rm}
\r_{mM+j}=\r_j,\qquad m=0,\ldots,n-1,\quad j=1,\ldots, M.
\ee
Then, for any bounded complex-valued function $f$ on $\Rr^d$,
\be\label{f-f}
\frac{1}{(nM)^2}\sum_{i,j=1}^{nM}f(\r_i-\r_j)=\frac{1}{M^2}\sum_{i,j=1}^Mf(\r_i-\r_j).
\ee
When $u$ is bounded, we apply this identity to $f=Vu_\Lambda/2$. Introducing
\be\label{omegarn}
\omega(\r)_N=\frac{V}{2N^2}\sum_{i,j=1}^N u_\Lambda(\r_i-\r_j)
=\frac{N-1}{N}\,\epsilon(\r)_N+\frac{Vu_\Lambda(\0)}{2N}
\ee
and
\be\label{omegan}
\omega_N=\min_{(\r)_N\subset\Lambda}\omega(\r)_N=\frac{N-1}{N}\,\epsilon_N+\frac{Vu_\Lambda(\0)}{2N}\,,
\ee
and recalling that $m|n$ for integers $m,n$ denotes that $m$ is a divisor of $n$, we obtain the following.

\begin{lemma}
\be\label{o-o}
\omega(\r)_M^n=\omega(\r)_M
\ee
and, hence,
\be\label{o-ineqs}
\omega_{N_1}\geq\omega_{N_2}\geq\omega_{N_3}\geq\cdots\quad {\rm if}\quad N_1|N_2|N_3|\cdots.
\ee
\end{lemma}
\emph{Proof.} Equation (\ref{o-o}) repeats (\ref{f-f}). Let $(\r)_{N_j}$ be a $N_j$-particle ground state. Then
\be
\omega_{N_{j+1}}\leq\omega(\r)_{N_{j}}^{N_{j+1}/N_{j}}=\omega(\r)_{N_{j}}=\omega_{N_{j}}.\quad \Box
\ee

\begin{proposition}\label{C-inf-omega} If $u$ is bounded,
\be\label{C=inf}
C_\Lambda[u]=\inf_N\omega_N.
\ee
\end{proposition}
\emph{Proof.}
\be\label{om-eps}
\omega_N-\epsilon_N=\frac{1}{N}\left(\frac{Vu_\Lambda(\0)}{2}-\epsilon_N\right)
\ee
goes to zero as $N$ increases, so the statement is
\be\label{inf-liminf}
\liminf_{N\to\infty}\omega_N=\inf_N\omega_N.
\ee
Now $\inf_N\omega_N\leq\liminf_{N\to\infty}\omega_N$; supposing a strict inequality, because of (\ref{o-ineqs}) there would be a sequence $N_j$ tending to infinity such that
\be
\omega_{N_j}\leq\omega_{N_1}<\liminf_{N\to\infty}\omega_N
\ee
contradicting the definition of liminf. $\Box$

A direct consequence of Equation~(\ref{C=inf}) is that
\be\label{super}
U_\Lambda(\r)_N\geq -\frac{u_\Lambda(\0)}{2}N+C_\Lambda[u]\frac{N^2}{V}
\ee
for \emph{all} $N$. Another consequence is as follows.

\begin{corollary}\label{propCinf} If $\omega(\r)_M=C_\Lambda[u]$, then $(\r)_M^n$ is a $nM$-particle ground state for all $n\geq 1$.
\end{corollary}
The simplest realization of this situation is provided by the penetrable sphere model and is presented in Section~\ref{sec-penetrable}. For a less trivial example see Ref.~\cite{Su4}.

\newsec{Infinite-density limit of the ground-state energy and the free energy per pair}\label{idens-limit}

\begin{proposition}\label{epsilonN} The sequence $\epsilon_N$ is convergent.
\end{proposition}
\emph{Proof.} We give two different proofs.\\
(i) The first proof works for $u$ bounded. In this case $\omega_N-\epsilon_N$ tends to zero with increasing $N$, therefore the convergence of $\epsilon_N$ is equivalent to
\be
\lim_{N\to\infty}\omega_N=\inf_N\omega_N=C_\Lambda[u].
\ee
Now $C_\Lambda[u]$ is the smallest accumulation point of the sequence $\omega_N$, and suppose there is another one, $\omega_0$. Let $\delta=\frac{1}{3}(\omega_0-C_\Lambda[u])$ and define two subsequences, $K_m$ and $L_n$ via the inequalities $\omega_{K_m}\leq C_\Lambda[u]+\delta$ and $\omega_{L_n}\geq\omega_0-\delta$; thus, $\omega_{L_n}-\omega_{K_m}\geq\delta$ for all $m,n$. Both subsequences are infinite and, because of Eq.~(\ref{o-ineqs}), $\ell K_1\in\{K_m\}$ for all integers $\ell\geq 1$. Therefore, for any $n$ there is an $m$ such that $0<L_n-K_m<K_1$ and, thus,
\be\label{LnKm}
E_0(L_n)\leq E_0(K_m)+\frac{1}{2}K_1(K_1-1+K_m)\|u_\Lambda\|_\infty\ .
\ee
The right member of this inequality is an upper bound to the energy of a $L_n$-particle configuration that one obtains from a $K_m$-particle ground state by adding $L_n-K_m$ particles. Dividing by $L_n^2/V$ we find
\be
\omega_{L_n}-\omega_{K_m}=O(L_n^{-1})
\ee
contradicting $\omega_{L_n}-\omega_{K_m}\geq\delta$.\\
(ii) The second proof works even if $u$ is unbounded. It is based on the monotonic increase of the ground state energy per pair,
\be\label{monoton}
\frac{E_0(N+1)}{N(N+1)}\geq\frac{E_0(N)}{N(N-1)};
\ee
see also Kiessling \cite{Kie}. If $u$ is integrable then $\epsilon_N=VE_0(N)/N(N-1)$ is bounded above by $v(\0)/2$, cf. Eq.~(\ref{min-av}), so its limit as $N$ goes to infinity exists,
\be\label{epsilon-limit}
\lim_{N\to\infty}\epsilon_N=\sup\epsilon_N=C_\Lambda[u].
\ee
As to Eq.~(\ref{monoton}), let $N\geq 2$ and $R$ be any sequence of $N+1$ points of $\Lambda$ (repetition allowed). Then
\be
U_\Lambda(R)=\sum_{(\x,\y)\subset R}u_\Lambda(\x-\y)=\frac{1}{N-1}\sum_{S\subset R, |S|=N}\sum_{(\x,\y)\subset S}u_\Lambda(\x-\y)=\frac{1}{N-1}\sum_{S\subset R, |S|=N}U_\Lambda(S),
\ee
because for any two-point subsequence $(\x,\y)\subset R$ we can choose $N-1$ different $N$-point subsequences $S\subset R$ containing $(\x,\y)$. Since the sum over $S$ has $N+1$ terms,
\be
\min_{R\subset\Lambda,|R|=N+1}U_\Lambda(R)
\geq\frac{N+1}{N-1}\min_{S\subset\Lambda, |S|=N}U_\Lambda(S)
\ee
which is just (\ref{monoton}). $\Box$

We have nowhere used the periodicity of the interaction, therefore Proposition~\ref{epsilonN} applies also to $u$ instead of $u_\Lambda$. It holds true also for unstable interactions, when $C_\Lambda[u]<0.$ If $u$ is bounded then Propositions~\ref{C-inf-omega} and \ref{epsilonN} yield
\be\label{sup-C-inf}
\sup\epsilon_N=C_\Lambda[u]=\inf\omega_N.
\ee

\begin{corollary}\label{coro-uniform}
If $u$ is bounded then the convergence of $\epsilon_{\lfloor\rho V\rfloor}$ and $\omega_{\lfloor\rho V\rfloor}$ to $C_\Lambda[u]$ as $\rho$ tends to infinity is uniform in $V$.
\end{corollary}
\emph{Proof.}
\be
0\leq \max\left\{\omega_{\lfloor\rho V\rfloor}-C_\Lambda[u],C_\Lambda[u]-\epsilon_{\lfloor\rho V\rfloor}\right\} \leq \omega_{\lfloor\rho V\rfloor}-\epsilon_{\lfloor\rho V\rfloor}\leq\frac{u_\Lambda(\0)}{2\rho}\leq\frac{u(\0)}{\rho}
\ee
if $\rho$ and $V$ are large enough. Here we used Eq.~(\ref{om-eps}) and superstability for the third and strong temperedness for the fourth inequality. $\Box$

The uniform convergence of $\epsilon_{\lfloor\rho V\rfloor}$ to $C_\Lambda[u]$ probably holds true also if $u$ diverges at the origin. If $\Lambda$ is a cube of side $L$ and $R$ is any $N$-point configuration then there is an $\x\in\Lambda$ such that dist$(\x,R)\geq L/(2N^{1/d})$. Using this fact, for a radial $u$ one can easily prove that
\be\label{epsN-N+1}
\epsilon_N\leq\epsilon_{N+1}\leq\epsilon_N+c\rho^{-1}u\left(\frac{1}{2\rho^{1/d}}\right)
\ee
if $\Lambda$ is large enough. Here $\rho=N/V$ and $c=1$ if $u(\x)$ is a monotonic function of $|\x|$ close to zero. The term $\rho^{-1}u(\rho^{-1/d}/2)$ is the analogue of $\rho^{-1}u(\0)$. Because $u$ is integrable, it tends to zero as $\rho$ goes to infinity. However, the right member of (\ref{epsN-N+1}) may not be an upper bound to $C_\Lambda[u]$ and, therefore, it cannot play the role of $\omega_N$ in the bounded case.

It is natural to ask how the free energy per pair behaves as the density tends to infinity. Let
\be\label{partition}
Z_{\Lambda,N}=\frac{1}{V^N}\int_{\Lambda^N}e^{-\beta U_\Lambda(\r)_N}\d(\r)_N\equiv e^{-\beta F_N(\beta)}.
\ee
For the free energy per pair, $f_N(\beta)$, we have
\be\label{trivibounds}
\epsilon_N\leq f_N(\beta)\equiv\frac{VF_N(\beta)}{N(N-1)}\leq \frac{v(\0)}{2}
\ee
where the upper bound results from Jensen's inequality. We are interested in the limit of $f_N(\beta)$ as $\beta$ or $N$ goes to infinity. The usual definition of the partition function is $V^N/N!$ times the expression (\ref{partition}). However, computing $-(V/\beta N(N-1))\ln Z_{\Lambda,N}$ with the modified definition or with (\ref{partition}) yields the same result for both limits. At positive temperatures, as $N$ goes to infinity, $f_N(\beta)-\epsilon_N$ remains positive only if the entropy is negative and of order $N^2$. This means that the level set
$$\{(\r)_N\in\Lambda^{N}: U_\Lambda(\r)_N\leq E_0(N)+N^2/\beta\}$$
should be of Lebesgue measure $\sim e^{-cN^2}$ with some $c>0$.
Under some natural conditions on the interaction we shall find the opposite result, an infinite-density limit of the free energy per pair that agrees with the limit of $\epsilon_N$, i.e., $C_\Lambda[u]$.

\begin{theorem}  Let $u$ satisfy one of the following conditions.\\
(i) $u$ is bounded and
\be\label{Lipb}
|u(\x)-u(\y)|\leq \left(|u(\x)|+|u(\y)|\right)\left(\frac{|\x-\y|}{r_0}\right)^\alpha
\ee
with some $r_0>0$ and $0<\alpha\leq 1$.\\
(ii) There exist constants $u_0>0$, $r_0>0$, $c>0$, $0<\zeta<d$ and $0<\alpha\leq 1$ such that
\be\label{alg}
u(\x)\geq u_0\left(\frac{r_0}{|\x|}\right)^{d-\zeta}\quad{\rm if}\quad |\x|\leq r_0
\ee
and
\be\label{Lipu}
|u(\x)-u(\y)|\leq c(|u(\x)|+|u(\y)|)\left(\frac{|\x-\y|}{\min\{|\x|,|\y|,r_0\}}\right)^\alpha.
\ee
Then
\be\label{beta-N-limits}
\lim_{\beta\to\infty}f_N(\beta)=\epsilon_N,\quad \lim_{N\to\infty}f_N(\beta)=\lim_{N\to\infty}\epsilon_N=C_\Lambda[u] \quad\mbox{all $\beta>0$}.
\ee
\end{theorem}
\emph{Proof.} (i) Using (\ref{Lipb}),
\be
|u_\Lambda(\x)-u_\Lambda(\y)|\leq 2\|u_\Lambda\|\left(\frac{|\x-\y|}{r_0}\right)^\alpha
\ee
where
\be
\|u_\Lambda\|=\sup_\x\sum_{\n\in\Zz^d}|u(\x+L\n)|.
\ee
Let $(\r)_N$ be a ground state configuration [hence, $U_\Lambda(\r)_N=E_0(N)$] and $(\r+\delta\r)_N=(\r_1+\delta\r_1,\ldots,\r_N+\delta\r_N)$ an arbitrary perturbation of it. Then
\be\label{U-Ugs}
U_\Lambda(\r+\delta\r)_N-E_0(N)\leq \|u_\Lambda\|\left(2\max_{i}\frac{|\delta\r_{i}|}{r_0}\right)^\alpha N^2\leq\beta^{-1}
\ee
if for all $i$
\be\label{deltabound}
|\delta\r_i|\leq\frac{r_0}{2} \left(\frac{1}{\beta\|u_\Lambda\|N^2}\right)^{\frac{1}{\alpha}}\equiv\frac{r_0}{2}\varepsilon_{\beta,N}.
\ee
Using Eqs.~(\ref{partition}), (\ref{trivibounds}), (\ref{U-Ugs}) and (\ref{deltabound}),
\be
1\geq e^{-\beta N(N-1)[f_N(\beta)-\epsilon_N]/V}\geq
\frac{c_d}{e}\left[\frac{(r_0\varepsilon_{\beta,N}/2)^d}{V}\right]^N
\ee
where $c_d$ is a dimension-dependent constant. Taking the logarithm, dividing by $\beta N(N-1)/V$ and letting either $\beta$ or $N$ go to infinity we find Eq.~(\ref{beta-N-limits}).\\
(ii) Let $(\r)_N$ be a ground state configuration and let $\r_{ij}=\r_i-\r_j$. Because of Eq.~(\ref{min-av}) and the lower boundedness of $u$,
$
u(\r_{ij}+L\n)\leq u_1N^2
$
holds with some $u_1>0$ for any $\n\in\Zz^d$. By decreasing, if necessary, $r_0$, (\ref{alg}) is valid also with $u_1$ replacing $u_0$ and implies
\be
|\r_{ij}+L\n|\geq r_0 N^{-\frac{2}{d-\zeta}}.
\ee
Let $\x_\n=\r_{ij}+\delta\r_{ij}+L\n$ and $\y_\n=\r_{ij}+L\n$ where $\delta\r_{ij}=\delta\r_{i}-\delta\r_{j}$, and choose
\be\label{dbound2}
|\delta\r_i|\leq\frac{r_0}{4} \left(\frac{1}{u_2\beta N^{4+\frac{2}{d-\zeta}}}\right)^{\frac{1}{\alpha}}
\ee
for every $i$, where $u_2=3cu_1$. If $u_2\beta N^4\geq 1$, this inequality implies
\be\label{deltabound'}
|\delta\r_{ij}|\leq\frac{r_0}{2} N^{-\frac{2}{d-\zeta}}.
\ee
Then
\be
\min\{|\x_\n|,|\y_\n|,r_0\}\geq \frac{r_0}{2} N^{-\frac{2}{d-\zeta}}
\ee
and by (\ref{Lipu}),
\be
|u(\x_\n)-u(\y_\n)|\leq c(|u(\x_\n)|+|u(\y_\n)|)\left(\frac{2|\delta\r_{ij}|}{r_0 N^{-\frac{2}{d-\zeta}}}\right)^\alpha.
\ee
When passing to $u_\Lambda$ we have to estimate $\sum_\n |u(\x_\n)|$ and $\sum_\n |u(\y_\n)|$. It is easily seen that only a single term can be large, the sum of the rest is of order 1 as $N$ goes to infinity. The possible large term is bounded by $u_1N^2$. Summation over $i,j$ brings in another factor $N^2$, and for $N$ or $\beta$ large enough we end up with a generous upper bound
\be\label{Unbound}
U_\Lambda(\r+\delta\r)_N-E_0(N)\leq u_2\left(4\max_{i}\frac{|\delta\r_{i}|}{r_0}\right)^\alpha N^{4+\frac{2}{d-\zeta}}\leq \beta^{-1}.
\ee
The rest of the proof is as in the bounded case. $\Box$

\emph{Remarks.} (i) For the validity of (\ref{beta-N-limits}) the continuity of $u$ was essential. We shall see that for the penetrable sphere model at least the first of Eqs.~(\ref{beta-N-limits}) fails. Condition (\ref{Lipb}) is a sort of strong H\"older-continuity (because $u(\x)$ decays as $|\x|$ grows) which guearantees the ordinary H\"older-continuity of $u_\Lambda$. In the case of a free boundary condition ordinary H\"older-continuity would suffice. (ii)~There is a gap between interactions that are bounded or diverge algebraically at the origin. Interactions with a logarithmic divergence, occurring in some cases between colloidal particles \cite{WP}, allow particles to be much closer to each other and, thus, the ground state configurations to be strongly inhomogeneous. One cannot exclude that instead of (\ref{deltabound}) one should impose $|\delta\r_{ij}|=O(e^{-cN})$. Then, one could still prove the first of Eqs.~(\ref{beta-N-limits}) but not the second one, because of an entropy $\propto -N^2$ at positive temperatures.

\newsec{Infinite-density ground state and the energy functional in finite volume}\label{sec-IDGS-dens-func}

\subsection{Bounded interactions}\label{sec-bounded}

In this section we show that it is possible to obtain $C_\Lambda[u]$ and a good approximation of high-density ground states via the minimization of an energy functional. For bounded interactions we can write $\omega(\r)_N$ in the form
\be\label{mupoint}
\omega(\r)_N=\frac{1}{2V}\int_{\Lambda^2}u_\Lambda(\x-\y)\d\mu_{(\r)_N}(\x)\d\mu_{(\r)_N}(\y)
\ee
where
\be\label{gspoint}
\mu_{(\r)_N}=\frac{V}{N}\sum_{j=1}^N\delta_{\r_j}
\ee
is a measure on $\Lambda$ of total weight $V$ (henceforth, a normalized measure); $\delta_\x$ is the Dirac delta at $\x$. We call $\mu_{(\r)_N}$ the measure associated with $(\r)_N$. The discrete measures of the form (\ref{gspoint}) -- sums of Dirac deltas with equal weights, sometimes called Dirac combs -- are dense among the normalized Borel measures relative to the weak topology, cf. Lemma A.1. Therefore, when $N$ goes to infinity and $R_N$ is a $N$-particle configuration, the sequence $\mu_{R_N}$ of associated measures can converge weakly to any normalized measure. We shall use this property to introduce the notion of an infinite-density ground state. Let $\calm$ denote the set of normalized Borel measures on $\Lambda$. A sequence $\mu_n\in\calm$ converges vaguely to a $\mu\in\calm$ if for every real continuous function $f$ of compact support
\be\label{muf}
\lim_{n\to\infty}\mu_n(f)=\mu(f),
\ee
where $\mu(f):=\int f\d\mu$. The convergence is in the weak sense if (\ref{muf}) holds for every bounded continuous $f$. Because $\Lambda$ is compact, every continuous function is bounded and of compact support, so the two types of convergences are the same. Henceforth $\mu_n\rightharpoonup\mu$ will denote that $\mu$ is the vague limit of the sequence $\mu_n$. For the reader's convenience, in the Appendix we show that any infinite sequence $\mu_n\in\calm$ has a weakly convergent subsequence (i.e., $\calm$ is compact in the weak topology), and any $\mu\in\calm$ is the weak limit of a sequence of discrete measures of the form (\ref{gspoint}).

\begin{definition}\label{def-IDGS}
$\mu\in\calm$ is an infinite-density ground state (IDGS) of $u_\Lambda$ if there is a sequence $\mu_{R_m}$ of measures associated with configurations $R_m$ such that $|R_m|$ goes to infinity with $m$ and
\be\label{IDGSdef}
\mu_{R_{m}}\rightharpoonup\mu,\qquad
\lim_{m\to\infty}\epsilon(R_{m})=C_\Lambda[u].
\ee
\end{definition}\label{defIDGS}
Note that $R_m$ may not be a sequence of ground state configurations.

\begin{proposition}\label{IDGSnonempty}
The set of IDGS's is nonempty.
\end{proposition}
\emph{Proof.} Given any sequence $\{R_N\}_1^\infty$ of $N$-particle ground states, $\epsilon(R_N)=\epsilon_{N}$ tends to $C_\Lambda[u]$, cf. Eq.~(\ref{epsilon-limit}). Moreover, the sequence $\mu_{R_N}$ of associated measures has at least one weak limit point which is, therefore, an IDGS. $\Box$

Motivated by Eq.~(\ref{mupoint}), we define an energy functional on $\calm$ by the equation
\be\label{Imu}
I[\mu]=\frac{1}{2V}\int_{\Lambda^2}u_\Lambda(\x-\y)\d\mu(\x)\d\mu(\y)=
\frac{1}{2V}\mu*\widetilde{\mu}(u_\Lambda).
\ee
Here we use the notations of Ref.~\cite{BF}: for a complex function $f$, $\widetilde{f}(\x):=\overline{f(-\x)}$, $\widetilde{\mu}(f):=\overline{\mu(\widetilde{f})}$. The measure $V^{-1}\mu*\widetilde{\mu}$ is the autocorrelation of $\mu$. It is defined on $\Lambda-\Lambda$ ($=\Lambda$ in the case of periodic boundary conditions when $\Lambda$ is a torus) by
\be
\mu*\widetilde{\mu}(f)=\int_{\Lambda-\Lambda}f(\z)\d(\mu*\widetilde{\mu})(\z) :=
\int_{\Lambda^2}f(\x-\y)\d\mu(\x)\d\mu(\y),
\ee
where $f$ is any continuous periodic function of period cell $\Lambda$. Thus,
\be
\mu*\widetilde{\mu}= \int_{\Lambda^2}\delta_{\x-\y}\d\mu(\x)\d\mu(\y) =\int_{\Lambda^2}\delta_{\x+\y}\d\mu(\x)\d\widetilde{\mu}(\y).
\ee
For general results about the autocorrelation and its Fourier transform see~\cite{BF,Hof}. Equation~(\ref{mupoint}) can be rewritten as $\omega(\r)_N=I[\mu_{(\r)_N}]$. $I[\mu]$ is meaningful also for unbounded interactions, but it can be finite only if $\mu$ is a continuous measure. For example, for the Lebesgue measure $\lambda$, $I[\lambda]=v(\0)/2$. In $I[\mu]$ one can recognize the interaction energy common to all mean-field theories, either quantum as the Hartree approximation~\cite{KN}, or classical as the mean-field theory of fluids~\cite{Lik,LLWL}. The difference in our use of it is that, while in mean-field theories $\mu$ is tacitly supposed to be absolutely continuous, here no \emph{a priory} assumption is made on it. Let
\be\label{I0}
I^0=\inf_{\mu\in\calm}I[\mu].
\ee
For bounded interactions
\be\label{I0Cu}
I^0\leq \inf_{N,(\r)_N} I[\mu_{(\r)_N}]=\inf_N\omega_N=C_\Lambda[u].
\ee
Before proving the opposite inequality in Theorem~\ref{bounded}, let us note that the correspondence between a finite configuration $(\r)_N$ and the associated measure (\ref{gspoint}) is not one-to-one. The configuration $(\r)_M^n$ defined by Eq.~(\ref{rm}) is different from $(\r)_M$ whose $n$-times repetition gives rise to it. This is in contrast with
\be\label{muident}
\mu_{(\r)_M^n}=\mu_{(\r)_M},
\ee
implied by Eq.~(\ref{o-o}). Thus, any discrete measure is associated with an infinity of finite configurations. This observation leads to the counterpart of Corollary~\ref{propCinf}.

\begin{proposition}\label{prop-piling}
Suppose that $\mu_{(\r)_M}$ is a minimizer of the energy functional. Then for any positive integer $n$, $(\r)_M^n$ is a $nM$-particle ground state of $u_\Lambda$.
\end{proposition}
\emph{Proof.}
Because of the identities (\ref{o-o}) and (\ref{muident}),
\be
\omega(\r)_M^n=I[\mu_{(\r)_M}]=I^0\leq I[\mu_{(\x)_{nM}}]=\omega(\x)_{nM}
\ee
and, hence,
\be
U_\Lambda(\r)_M^n\leq U_\Lambda(\x)_{nM}
\ee
for any $nM$-particle configuration $(\x)_{nM}$. $\Box$

\noindent
In this way, if a minimizer of $I[\mu]$ happens to be a discrete measure concentrated e.g. on the vertices of a lattice inside $\Lambda$, then the particles pile up on the lattice sites already in finite-density ground states.

For a $\Lambda$-periodic bounded Borel function $f$ and a $\mu\in\calm$, let
\be\label{If}
I_f[\mu]=\frac{1}{2V}\mu*\widetilde{\mu}(f).
\ee
$I[\mu]\equiv I_{u_\Lambda}[\mu]$ in this notation.

\begin{lemma}\label{lem-weaklim} If $u$ is bounded then $I[\mu]$ is continuous in the weak topology, i.e., $I[\mu_n]\to I[\mu]$ if $\mu_n\rightharpoonup\mu$.
\end{lemma}
\emph{Proof.} Let first $u$ be continuous. If $\mu_n\rightharpoonup\mu$ then $\mu_n*\widetilde{\mu_n}\rightharpoonup \mu*\widetilde{\mu}$, therefore
\be
\mu_n*\widetilde{\mu_n}(f)\rightarrow \mu*\widetilde{\mu}(f)
\ee
if $f$ is continuous on $\Lambda$. Applying this to $f=u_\Lambda$ the result follows. Suppose now that $u$ is bounded. For any $\varepsilon>0$ we can choose a real continuous $f$ on $\Lambda$ and an integer $m$ (depending also on $f$) such that $\|u_\Lambda-f\|_\infty<\varepsilon$ and $|I_{f}[\mu]-I_{f}[\mu_n]|<\varepsilon$ for $n\geq m$. Then $|I_{u_\Lambda}[\mu]-I_{u_\Lambda}[\mu_n]|<3\varepsilon$ for $n\geq m$ and, hence,
\be
\lim_{n\to\infty} |I_{u_\Lambda}[\mu]-I_{u_\Lambda}[\mu_n]|\leq 3\varepsilon.
\ee
This being true for every $\varepsilon>0$, the above limit actually vanishes.  $\Box$

\begin{theorem}\label{bounded} If $u$ is bounded, then
\be\label{I0Clam}
I^0=C_\Lambda[u],
\ee
the infimum $I^0$ is attained, and $I[\mu]=I^0$ if and only if $\mu$ is an IDGS.
\end{theorem}
\emph{Remark.} Because of Eq.~(\ref{C=inf}), Equation (\ref{I0Clam}) states that the infimum of $I[\mu]$ over all $\mu\in\calm$ is the same as the infimum over discrete measures associated with finite-density ground states. This does not mean that the IDGS cannot be a continuous measure, see the forthcoming sections.

\noindent
\emph{Proof.} Choose $\mu^m\in\calm$ such that $I[\mu^m]$ tends to $I^0$ as $m$ goes to infinity ($\mu^m$ may be the same for each $m$). Because the set of discrete measures (\ref{gspoint}) is dense in $\calm$ and $I$ is continuous, $\mu^m$ can be chosen to be of the form (\ref{gspoint}); hence, $I[\mu^m]\geq C_\Lambda[u]$. If $\mu$ is any weak limit point of $\mu^m$ then by the continuity of $I$,
\be\label{double}
C_\Lambda[u]\geq I^0=\lim_{m\to\infty}I[\mu^m]=I[\mu]\geq C_\Lambda[u].
\ee
This proves (\ref{I0Clam}) and that $\mu$ is an IDGS. Starting with an IDGS $\mu$ and choosing $\mu^m=\mu_{R_m}$ where $R_m$ is the defining sequence (\ref{IDGSdef}), Eq.~(\ref{double}) shows that $\mu$ is a minimizer of $I$. $\Box$

\noindent
\emph{Remark.} We have found that weak limits of $N$-particle ground states are minimizers of $I$, but have not proved that every minimizer$\,=\,$IDGS can be obtained as such a limit.

\subsection{Unbounded interactions}\label{sec-unbounded}

We restrict the discussion to pair potentials such that $u(\x)\to\infty$ as $\x\to 0$ and $u$ is bounded otherwise. Definition~\ref{defIDGS} and Proposition~\ref{IDGSnonempty} are valid in this case. Because the notion of an IDGS is independent of the energy functional, the divergence of $I[\mu]$ on point measures does not allow to directly conclude that IDGS's are continuous measures. Below we present a proof which is independent of $I$.

\begin{proposition}\label{propunbounded} Let $u$ be integrable and $u(\x)\to\infty$ as $\x\to \0$. Then any IDGS $\mu$ is purely continuous.
\end{proposition}
\emph{Proof.} Take any $\x_0\in\supp\mu$. Let $R_m$ be the defining sequence of $\mu$, cf. Definition~\ref{def-IDGS}. Because $u_\Lambda(\0)=\infty$, all points of $R_m$ are distinct. Choose any $\varepsilon>0$ and let $C$ be an open ball of diameter $\varepsilon$ centered at $\x_0$. By approximating the characteristic function of $C$ with continuous functions one can see that
$$\lim_{m\to\infty}\mu_{R_m}(C)=\mu(C).$$
It follows that for $m$ large enough $\mu_{R_m}(C)\geq \frac{1}{2}\mu(C)$ and, because $\mu_{R_m}(\{\x\})=V/|R_m|$ for $\x\in R_m$,
\be
|R_m\cap C|\geq\frac{\mu(C)|R_m|}{2V}.
\ee
As a consequence,
\be
U_\Lambda(R_m\cap C)\geq \frac{\mu(C)|R_m|}{4V}\left(\frac{\mu(C)|R_m|}{2V}-1\right)\inf_{|\x|<\varepsilon}u_\Lambda(\x)
\ee
and
\be
U_\Lambda(R_m)\geq |R_m|^2\left\{\left[\frac{\mu(C)^2}{8V^2}-\frac{\mu(C)}{4V|R_m|}\right]\inf_{|\x|<\varepsilon}u_\Lambda(\x)-\frac{\sup u_\Lambda^-}{2}\right\}
\ee
where $u_\Lambda^-(\x)=-\min\{u_\Lambda(\x), 0\}$. Because $u$ is bounded below and strongly tempered, $\sup u_\Lambda^-$ is finite. Dividing by $|R_m|(|R_m|-1)/V$ and letting $m$ tend to infinity,
\be
C_\Lambda[u]\geq \frac{\mu(C)^2}{8V}\inf_{|\x|<\varepsilon}u_\Lambda(\x)-\frac{V\sup u_\Lambda^-}{2}
\ee
or
\be
\mu(C)^2
\leq \frac{4V(2C_\Lambda[u]+V\sup u_\Lambda^-)}{\inf_{|\x|<\varepsilon}u_\Lambda(\x)}.
\ee
The upper bound vanishes as $\varepsilon$ goes to zero, proving that $\mu(\{\x_0\})=~0$. $\Box$

The extension of Theorem~\ref{bounded} to unbounded interactions is not immediate because IDGS's are weak limits of discrete measures which make $I[\mu]$ diverge. We therefore introduce auxiliary functionals which can give finite values on point measures. For a positive $K$ let
\be
\widetilde{I}_{K}[\mu]=\frac{1}{2V}\int_{\{(\x,\y):u_\Lambda(\x-\y)\leq K\}}u_\Lambda(\x-\y)\d\mu(\x)\d\mu(\y)
\ee
and
\be
\widetilde{I}[\mu]=\lim_{K\to\infty}\widetilde{I}_{K}[\mu]=\sup_K \widetilde{I}_{K}[\mu].
\ee
If $u$ is bounded, $\widetilde{I}[\mu]\equiv I[\mu]$. If $u$ is unbounded, $\widetilde{I}[\mu]\leq I[\mu]$. In contrast to $I$, $\widetilde{I}$ can be finite on point measures if $u(\x)$ tends to infinity when $\x\to 0$ but $u$ is bounded outside the origin, in which case we also have
\be
\widetilde{I}[\mu]=\frac{1}{2V}\lim_{\varepsilon\to 0}\int_{d_\Lambda(\x,\y)\geq\varepsilon}u_\Lambda(\x-\y)\d\mu(\x)\d\mu(\y)
\ee
where $d_\Lambda(\x,\y)=\min_{\n\in\Zz^d}|\x-\y+L\n|$. The usefulness of $\widetilde{I}$ is based on the easily verifiable fact that $\widetilde{I}[\mu_{(\r)_N}]=\epsilon(\r)_N$, and on $\widetilde{I}[\mu]=I[\mu]$ (maybe infinite) for continuous measures.

\begin{lemma}\label{lemcont} If $\mu$ is continuous then $I[\mu]=\widetilde{I}[\mu]\leq\infty$.
\end{lemma}
\emph{Proof.} Let
\be
g_n=\frac{1}{2V}(\mu\times\mu)(\{(\x,\y)\in\Lambda\times\Lambda:u_\Lambda(\x-\y)>n\}),
\ee
a monotone decreasing sequence. By the definition of the Lebesgue integral,
\be
I[\mu]=\lim_{K\to\infty}\left[\widetilde{I}_K[\mu]+Kg_K\right]=\widetilde{I}[\mu]+\lim_{K\to\infty}Kg_K,
\ee
implying that the limit exists ($\leq\infty$). If $\mu$ has a point part then
\be
2V\lim_{K\to\infty}g_K=(\mu\times\mu)\left(\bigcap_{n=1}^\infty\{u_\Lambda>n\}\right)=
\sum_{\x\in\Lambda}\mu(\{\x\})^2>0,
\ee
therefore $I[\mu]=\infty$, but $\widetilde{I}[\mu]$ is finite if e.g. $\mu$ is concentrated on a finite number of points. Suppose that $\mu$ is continuous, then $\lim g_K=0$. If $\lim Kg_{K}=0$, we have the claimed equality. If $\lim Kg_{K}\geq c>0$, introduce
\be
\widetilde{I}_K^+[\mu]=\widetilde{I}_K[\mu]-\widetilde{I}_0[\mu]=\widetilde{I}_K[\mu]+I_{u_\Lambda^-}[\mu].
\ee
Because $u_\Lambda$ is bounded below, the second term is finite. Now
\be
\widetilde{I}_K^+[\mu]\geq \sum_{n=1}^{K-1}n(g_n-g_{n+1})=\sum_{n=1}^{K}(g_n-g_K)\geq \sum_{n=1}^{K'}(g_n-g_K)
\ee
for any $K'\leq K$, therefore
\be
\lim_{K\to\infty}\widetilde{I}_K^+[\mu]\geq \sum_{n=1}^{K'}g_n
\ee
for any $K'$. It follows that
\be
\lim_{K\to\infty}\widetilde{I}_K^+[\mu]\geq \sum_{n=1}^{\infty}g_n=\infty
\ee
because $g_n>c/2n$ for $n$ large enough. Thus, in this case $I[\mu]=\widetilde{I}[\mu]=\infty$. $\Box$

\begin{lemma}\label{lemItilde} If $\mu$ is an IDGS then $\widetilde{I}[\mu]\leq C_\Lambda[u]$.
\end{lemma}
\emph{Proof.} Let $R_m$ be the defining sequence of $\mu$. By Lemma~\ref{lem-weaklim} and Theorem~\ref{bounded},
\be
\widetilde{I}_K[\mu]=\lim_{m\to\infty}\widetilde{I}_K[\mu_{R_m}] \leq\lim_{m\to\infty}\widetilde{I}[\mu_{R_m}]=\lim_{m\to\infty}\epsilon(R_m)=C_\Lambda[u].
\ee
Taking the supremum over $K$ gives the result. $\Box$

\begin{lemma}\label{CI0} $C_\Lambda[u]\leq I^0$.
\end{lemma}
\emph{Proof.} Define
\be\label{uk}
u_\Lambda^K(\x)=\min\{u_\Lambda(\x),K\}.
\ee
For any $N$ there exists some $K_N$ such that $R_N$ is a $N$-particle ground state of $U_\Lambda$ if and only if it is also a $N$-particle ground state of $U_\Lambda^K$ for any $K\geq K_N$, and
\be\label{UKU}
E_0^K(N)=U_\Lambda^K(R_N)=U_\Lambda(R_N)=E_0(N).
\ee
For example, $K_N=E_0(N)+(N^2/2)\sup u_\Lambda^-$ is a possible choice. Indeed,
$u_\Lambda^K\leq u_\Lambda$, therefore $E_0^K(N)\leq E_0(N)$. It follows that if $K\geq K_N$ and $R_N$ is a $N$-particle ground state of either $U_\Lambda$ or $U_\Lambda^K$, then
$$u_\Lambda^K(\x-\y)=u_\Lambda(\x-\y)\leq K$$
for every pair $\{\x,\y\}\subset R_N$, otherwise
$$U_\Lambda(R_N)\geq U_\Lambda^K(R_N)>E_0(N).$$
This means, however, that $R_N$ is a common ground state of $U_\Lambda$ and $U_\Lambda^K$, and (\ref{UKU}) holds true. With the above $K_N$ and $R_N$, in an obvious notation, cf. (\ref{If}),
\be
I^0_{u_\Lambda^{K_N}}=C_\Lambda[u^{K_N}]\geq\epsilon_N^{K_N}=\epsilon_N,
\ee
the three relations holding due to Eqs.~(\ref{I0Clam}), (\ref{epsilon-limit}) and (\ref{UKU}), respectively. Letting $N$ tend to infinity, or taking the supremum over $N$,
\be
I^0\equiv I^0_{u_\Lambda}\geq\lim_{N\to\infty}I^0_{u_\Lambda^{K_N}}\geq C_\Lambda[u],
\ee
where the first inequality follows from Eq.~(\ref{uk}). $\Box$

Combining Proposition~\ref{propunbounded} and Lemmas~\ref{lemcont}, \ref{lemItilde} and \ref{CI0}, for any IDGS $\mu$ we find
\be
\widetilde{I}[\mu]=C_\Lambda[u]=I^0=I[\mu]
\ee
and, therefore, the following result.

\begin{theorem}\label{unbounded} If $u$ is integrable, bounded outside the origin, and $u(\x)\to\infty$ as $\x\to \0$, then $I^0=C_\Lambda[u]$, the infimum is attained, and any IDGS is continuous and minimizes $I$.
\end{theorem}

\noindent
\emph{Remark to Sections~\ref{sec-bounded} and \ref{sec-unbounded}.} In the proof of Theorems~\ref{bounded} and \ref{unbounded} we have not used that $u_\Lambda$ was a periodized interaction. Therefore, the theorems are valid in the case when $\Lambda$ is any compact Lebesgue-measurable set and the interaction is the original, non-periodic $u$.

\subsection{Fourier representation of the energy functional}\label{sec-Fourier}

Most of the results on IDGS's will be found by writing $I[\mu]$ in Fourier representation. This is obtained by substituting the expansion (\ref{exp}) into Eq.~(\ref{Imu}) and integrating term by term:
\be\label{nonneg}
I[\mu]=\frac{1}{2}\left[v(\0)+ \sum_{\0\neq\k\in\Lambda^*}v(\k)|\widehat{\mu}(\k)|^2\right]=\frac{1}{2V} \widehat{\mu*\widetilde{\mu}}(v).
\ee
Here
\be\label{muhat}
\widehat{\mu}(\k)=\frac{1}{V}\int_\Lambda e^{-i\k\cdot\x}\d\mu(\x),
\ee
so that $|\widehat{\mu}(\k)|\leq \widehat{\mu}(\0)=1$, and
\be\label{autocorr-Fourier}
\frac{1}{V}\widehat{\mu*\widetilde{\mu}}=\sum_{\k\in\Lambda^*}|\widehat{\mu}(\k)|^2\delta_\k.
\ee
We do not need the equality (\ref{nonneg}) to hold for all $\mu\in\calm$. We are interested in IDGS's, and $I[\mu]=C_\Lambda[u]\leq v(\0)/2$ for any of them. It suffices, therefore, that the Fourier expansion (\ref{nonneg}) converges to $I[\mu]$, whenever $I[\mu]\leq v(\0)/2$. In the case $v\geq 0$, $I[\mu]\leq v(\0)/2$ holds (with equality) if and only if $\mu$ is an IDGS; later on, we will discuss this case in detail. Let
\be
v^+(\k)=\max\{0,v(\k)\},\quad v^-(\k)=-\min\{0,v(\k)\}.
\ee
If $v^-\neq 0$, then the IDGS's are among those $\mu$ making the right member of Eq.~(\ref{nonneg}) absolutely convergent and satisfying
\be
\sum_{\0\neq\k\in\Lambda^*}v^-(\k)|\widehat{\mu}(\k)|^2-\sum_{\0\neq\k\in\Lambda^*}v^+(\k)|\widehat{\mu}(\k)|^2\geq 0.
\ee
For a stable interaction the left member of this inequality cannot exceed $v(\0)$, cf. Proposition~\ref{prop-stab}. The verity of (\ref{nonneg}) in this case can be seen as follows. Given a $t>0$, consider the Gaussian
\be
G^t(\x)=(4\pi t)^{-d/2}e^{-|\x|^2/4t}
\ee
and its periodization, called the periodic heat kernel,
\be
G^t_\Lambda(\x)=\sum_{\n\in\Zz^d}G^t(\x+L\n)=\frac{1}{V}\sum_{\k\in\Lambda^*}e^{-t|\k|^2}e^{i\k\cdot\x},
\ee
where the right member is obtained by the Poisson summation formula (or the Fourier expansion of the periodic function on the left). If $\mu\in\calm$, then $G^t_\Lambda*\mu\in\calm$ as well. [$f*\mu(\x):=\int f(\x-\y)\d\mu(\y)$.] Moreover, $G^t_\Lambda*\mu$ is absolutely continuous, and its Radon-Nikodym derivative
\be
\phi(\x)=\sum_{\k\in\Lambda^*} e^{-t|\k|^2}\widehat{\mu}(\k)e^{i\k\cdot\x}
\ee
is a real entire function of each component of $\x$. These properties guarantee that the energy functional evaluated on $G^t_\Lambda*\mu$ satisfies (\ref{nonneg}),
\be\label{IGt}
I[G^t_\Lambda*\mu]=\frac{1}{2}\sum_{\k\in\Lambda^*}v(\k)e^{-2t|\k|^2}|\widehat{\mu}(\k)|^2.
\ee
Because the left member is finite and the right member is absolutely convergent at $t=0$, the regularity of Abel summability implies that Eq.~(\ref{IGt}) survives the $t\to 0$ limit and yields (\ref{nonneg}).

The form (\ref{nonneg}) of $I[\mu]$ reveals an important property of the minimizers, hidden in the defining equation (\ref{Imu}). For stable interactions $0\leq I^0\leq v(\0)/2$ but, because $\Lambda^*$ is a set of density $\sim L^d$, for most $\mu\in\calm$ one finds $I[\mu]$ of the order of some power of $L$. For example,
\be
I[V\delta_\x]=\frac{1}{2}\sum_{\k\in\Lambda^*}v(\k)=Vu_\Lambda(\0)/2.
\ee
Thus, one may expect that when $\Lambda$ increases, the minimizers are among $\mu$'s such that for some volume-independent constant $c$ and for any $K>0$
\be
\left|\supp\widehat{\mu}\bigcap\{\k\in\Lambda^*:|\k|\leq K\}\right|\leq c K^d.
\ee
In Section~\ref{sec-stat} we shall meet some special sets of this kind, namely, additive subgroups of $\Lambda^*$. These will be shown to support the Fourier transform of the stationary points of $I$. When $v\geq 0$ or $u(\x)\geq \mbox{const}\times e^{-\alpha|\x|^2}$, at least the uniform local boundedness as $\Lambda\to\infty$ of both the autocorrelation and its Fourier transform (\ref{autocorr-Fourier}) is easily seen. We return to this question in Section~\ref{sec-IDGS-infinite}.

\newsec{Stability conditions for pair potentials}\label{sec-stability}

From Theorems~\ref{bounded} and \ref{unbounded} we can deduce stability conditions on $u$. If $u$ is nonnegative then it is obviously stable; if $u\geq 0$ and is integrable then its Fourier transform $v$ is a positive definite function (a function of positive type)~\cite{RSi}, therefore $v(\0)-|v(\k)|\geq 0$ for all $\k\in\Rr^d$. Due to Theorems~\ref{bounded} and \ref{unbounded}, similar conditions can be obtained without assuming $u\geq 0$.

\begin{proposition}\label{prop-stab} Let $u$ be an integrable pair potential with $u(-\x)=u(\x)$. Then $I^0=I^0(\Lambda)\geq 0$ for any cube $\Lambda$ is sufficient and necessary for stability, and  $I^0(\Lambda)>0$ for all $\Lambda$ is sufficient and necessary for superstability. Some particular necessary conditions for the stability of $u$ are as follows.

\noindent(i)
\be
2v(\0)+v(\k)\geq 0,\quad\mbox{all $\k\in\Rr^d$.}
\ee
(ii) If $u$ is radial and $v(\k)\leq 0$ for $|\k|\geq k_0$ then
\be\label{nnn}
v(\0)+n_d\min_{|\k|\geq k_0}v(\k)\geq 0
\ee
where $n_d$ is the number of nearest neighbors in a $d$-dimensional close-packed lattice ($n_d=6,12,24,40,\ldots$ for $d=2,3,4,5,\dots$).

\noindent(iii) If $u$ is bounded, radial and $u(\r)\leq 0$ for $|\r|\geq r_0$ then
\be\label{condonu}
u(\0)+n_d\min_{|\r|\geq r_0}u(\r)\geq 0.
\ee
\end{proposition}

\noindent
\emph{Remark.} Ruelle's example of a catastrophic potential (\cite{Rue}, Section 3.2.3) violates the inequality (\ref{condonu}).

\noindent
\emph{Proof.}
Because for $u$ bounded $I^0=\inf_{N,(\r)_N} \omega(\r)_N$, the condition $I^0\geq 0$ is part of Proposition 3.2.2 of Ruelle~\cite{Rue}. The role of the sign of $I^0$ for $u$ unbounded is also clear from $I^0=C_\Lambda[u]$. Now $v(-\k)=v(\k)$ real, and assertions (i)-(ii) are nontrivial if $v$ takes on negative values.

\noindent(i) Suppose that $v(\k)<0$ for some $\k\neq \0$ and consider the measure
\be
\d\mu_\k(\x)=(1+\cos\k\cdot\x)\d\x.
\ee
The form (\ref{nonneg}) of $I[\mu]$ shows (cf. also Proposition~\ref{prop-vgeq0}) that $I[\mu_\k]=v(\0)/2+v(\k)/4$ which must be nonnegative if $u$ is stable.

\noindent(ii) Let $\min_{|\k|\geq k_0}v(\k)=v(\k_m)$. Choose $\Lambda$ and a Bravais lattice $B$ (lattice, for mathematicians) such that the reciprocal lattice $B^*$ (dual lattice, defined by $\k\cdot\r\in2\pi\Zz$ for $\k\in B^*$ and $\r\in B$) is close-packed, $B^*\subset\Lambda^*$ and the nearest neighbor distance of $B^*$ is $|\k_m|$. Then, for $(\r)_N=B\cap\Lambda$, by a simple computation
\be\label{stability-cond-v}
I[\mu_{B}]\leq \frac{1}{2}[v(\0)-n_d|v(\k_m)|],
\ee
and the left member is nonnegative if $u$ is stable. Here $I[\mu_{B}]=I[\mu_{B\cap\Lambda}]$ which is independent of $\Lambda$ provided that $\Lambda$ is a period cell for $B$, see also Eq.~(\ref{I[muB]}).

\noindent(iii) Condition (\ref{condonu}) is the dual of (\ref{nnn}), obtained from the Poisson summation formula
\be
\sum_{\r\in B}u(\r)=\rho(B)\sum_{\k\in B^*}v(\k)=2\rho(B)I[\mu_B].
\ee
Here $\rho(B)$ denotes the density of $B$. Let $\min_{|\r|\geq r_0}u(\r)=u(\r_m)$. Choose $B$ to be close-packed with nearest-neighbor distance $|\r_m|$. Then, in case of stability,
\be
0\leq I[\mu_B]=\frac{1}{2\rho(B)}\sum_{\r\in B}u(\r)\leq\frac{1}{2\rho(B)}[u(\0)-n_d|u(\r_m)|]
\ee
which was to be shown. $\Box$

\newsec{Stationary points of the energy functional}\label{sec-stat}

The natural candidates for IDGS's are the stationary points of $I[\mu]$. To motivate the definition of stationary points, we first consider absolutely continuous normalized measures,
\be\label{muac0}
\d\mu(\x)=f^2(\x)\d\x,
\ee
where $f$ is a real function. We have to minimize $I[\mu]$ with respect to $f$ under the condition $\int_\Lambda f^2(\x)\d\x=V$. Via functional derivation of $I[\mu]$ we obtain
\be\label{muac}
f(\x)\left[\int_\Lambda u_\Lambda(\x-\y)f^2(\y)\d\y - c\right]=0.
\ee
Here $c$ is a Lagrange multiplier. Another way to write this equation is
\be\label{muacbis}
\int_\Lambda u_\Lambda(\x-\y)f^2(\y)\d\y= c\quad \mbox{if $f(\x)\neq 0$.}
\ee
Any absolutely continuous $\mu$ with a density $f^2$ solving Eq.~(\ref{muac}) is a stationary point of the energy functional. Because with such an $f$
\be
I[\mu]=\frac{1}{2V}\int_{\Lambda^2} u_\Lambda(\x-\y)f^2(\y)f^2(\x)\d\y\d\x=\frac{c}{2},
\ee
among the solutions of (\ref{muac}) one has to choose the one giving the smallest constant $c$. A solution of (\ref{muac}) is $f^2(\x)\equiv 1$ which defines the Lebesgue measure and yields $c=v(\0)$ and, hence, an already familiar upper bound $I^0\leq v(\0)/2$, cf. Eqs.~(\ref{Cv0}), (\ref{I0Cu}). We shall see that the Lebesgue measure is the only absolutely continuous measure among the stationary points of the functional $I$. By generalizing (\ref{muacbis}) we arrive at the following definition.

\begin{definition} A $\mu\in\calm$ is a stationary point of $I_f$, $\mu\in\calm^{\rm st}(f)$, if
\be\label{statf}
(f*\mu)(\x)=\int_\Lambda f(\x-\y)\d\mu(\y)=\mbox{{\rm constant}$\quad$ for $\x\in\supp\mu$.}
\ee
$\mu\in\calm$ is a stationary point (of $I$), $\mu\in\calm^{\rm st}$, if it is a stationary point of $I_f$ for every continuous function $f$.
\end{definition}
\emph{Remarks.} (i) One could ask $f*\mu$ to be constant only $\mu$-almost everywhere, but the above choice is sufficient and more convenient for our purposes. (ii) The name is justified by the following property. Let $\mu\in\calmst(f)$ and let $\nu\in\calm$, $\nu\ll\mu$. Then
\be
\int_\Lambda (f*\mu)\d\mu=\int_\Lambda (f*\mu)\d\nu,
\ee
and therefore
\be
I_f[(1-\veps)\mu+\veps\nu]-I_f[\mu]=\veps^2\left(I_f[\nu]-I_f[\mu]\right)
\ee
for any $0\leq\veps\leq 1$. So $\mu$ may not minimize $I_f$, not even among the measures absolutely continuous with respect to it, but if we perturb $\mu$ with a $\nu\ll\mu$ in the order of $\veps$, $I_f$ changes only in the order of $\veps^2$.

$\calm^{\rm st}\subset\calm^{\rm st}(f)$, and we shall give two examples showing that the first can be a proper subset of the second. $\calm^{\rm st}$ is nonempty, the Lebesgue measure is a stationary point. Intuitively, it appears to be clear that $\mu$ is a stationary point if it is uniform on its support, and $\supp\mu-\x\equiv\{\y-\x|\y\in\supp\mu\}$ is the same for all $\x\in\supp\mu$. The theorem below provides the proof.

For $\t\in\Lambda$ and a function $g$ on $\Lambda$ define $T_\t g$ by
\be
T_\t g(\x)=g(\x-\t),
\ee
and for $\mu\in\calm$, $T_\t\mu\in\calm$ by
\be
T_\t\mu(g)=\mu(T_{-\t}g).
\ee
Then for Borel sets $D\subset\Lambda$,
\be
T_\t\mu(D)=T_\t\mu(1_D)=\mu(D-\t).
\ee

\begin{lemma} Any translate of a stationary point of $I_f$ is a stationary point of $I_f$.
\end{lemma}
\emph{Proof.} Suppose $\mu\in\calmst(f)$.
Let $A=\supp\mu$, and $(f*\mu)(\x)\equiv c$ for $\x\in A$. Then $(f*T_\t\mu)(\x)\equiv c$ for $\x\in A+\t$ and, because $A+\t=\supp T_\t\mu$, we conclude that $T_\t\mu\in\calmst(f)$. $\Box$

The set $\Lambda=L\Tt^d$ is an additive group, $\Lambda-\Lambda=\Lambda$. A closed subset $G$ of $\Lambda$ is a closed additive subgroup if $G-G\subset G$. A Haar measure $\mu$ on $G$ is a nonzero measure of support $G$ which is invariant under shifts with elements of $G$; that is, $T_\t\mu=\mu$ for every $\t\in G$, $\mu$ is uniformly distributed on $G$. The normalized Haar measure is unique.

\begin{theorem}\label{stationary} A $\mu\in\calm$ is a stationary point if and only if it is a translate of the normalized Haar measure on a closed additive subgroup of $\Lambda$.
\end{theorem}
\emph{Examples.} Some simple closed additive subgroups with their normalized Haar measures are
\be\label{onepoint}
\{\0\}\qquad \mbox{with} \qquad V\delta_\0,
\ee
\be\label{uniform}
\Lambda \qquad \mbox{with} \qquad\lambda_d
\ee
and
\be
\left(B^{(n)}\cap\Lambda^{(n)}\right)\times\Lambda^{(d-n)}
\ee
with
\be\label{lattice}
\mu_{B^{(n)}\cap\Lambda}\times\lambda_{d-n}\equiv\frac{\lambda_n(\Lambda^{(n)})}{|B^{(n)}\cap\Lambda^{(n)}|}\sum_{\r\in B^{(n)}\cap\Lambda^{(n)}}\delta^{(n)}_\r\times\lambda_{d-n}
\ee
where $n=1,\ldots,d$, $\lambda_{n}$ and $\delta^{(n)}_\r$ are the $n$ dimensional Lebesgue and Dirac measures, respectively, $\Lambda^{(n)}$ is the projection of $\Lambda$ to the first $n$ coordinates and $B^{(n)}$ is a $n$-dimensional Bravais lattice commensurate with $\Lambda^{(n)}$. Any Bravais lattice can appear as a stationary point if we replace $\Lambda=L\Tt^d$ by a suitable parallelepiped.

In the following lemma we give a full description of closed additive subgroups of $\Lambda$ and the Haar measures on them. Related results can be found in Ref.~\cite{BF}. Recall that $\Lambda^*=(2\pi/L)\Zz^d$ and $V=L^d$.

\begin{lemma}\label{lem-Haar} Let $G$ be a closed additive subgroup of $\Lambda$ with normalized Haar measure $\mu$ (i.e., $\mu(G)=V$), and let
\be
G^*=\{\k\in\Lambda^*|\k\cdot\x\in 2\pi\Zz,\ \mbox{all $\x\in G$}\}.
\ee
Then $G^*$ is an additive subgroup of $\Lambda^*$, and $\widehat{\mu}$ is a Haar measure (the counting measure) on $G^*$,
\be\label{muhatHaar}
\widehat{\mu}(\k)=\sum_{\K\in G^*}\delta_{\k,\K}.
\ee
Conversely, if $G^*$ is an additive subgroup of $\Lambda^*$ with the counting measure $\widehat{\mu}$ on it, then
\be\label{Gconst}
G=\{\x\in\Lambda|\k\cdot\x\in 2\pi\Zz,\ \mbox{all $\k\in G^*$}\}
\ee
is a closed additive subgroup of $\Lambda$, and the measure $\mu$ defined by $\widehat{\mu}$ as its Fourier transform is a Haar measure on $G$ with $\mu(G)=V$.
\end{lemma}
\emph{Proof.} The (closed) subgroup property is obvious in both cases. Let now $G$ be a closed additive subgroup of $\Lambda$, either given initially or constructed as in (\ref{Gconst}), $G^*$ the dual of $G$, and $\mu$ the normalized Haar measure on $G$. We show that $\widehat{\mu}$ is just (\ref{muhatHaar}). For any $\k\in\Lambda^*$ and $\y\in G$,
\bea
\widehat{\mu}(\k)&=&\frac{1}{V}\int_G e^{-i\k\cdot\x}\d\mu(\x)=\frac{1}{V}\int_G e^{-i\k\cdot\x}\d\mu(\x+\y)\nonumber\\
&=&\frac{1}{V}\int_G e^{-i\k\cdot(\x-\y)}\d\mu(\x) = e^{i\k\cdot\y}\widehat{\mu}(\k).
\eea
Then either $\widehat{\mu}(\k)=0$ or $\k\cdot\y\in 2\pi\Zz$ for all $\y\in G$, that is, $\k\in G^*$. In the second case $\widehat{\mu}(\k)=1$, proving the claim. $\Box$

Thus, we have obtained a simple characterization of closed additive subgroups of the torus $\Lambda$ as the dual sets, in the sense of Eq.~(\ref{Gconst}), of intersections of linear subspaces of $\Rr^d$ with the lattice $\Lambda^*$. Indeed, each additive subgroup of $\Lambda^*$ is such an intersection. We now see that $\Lambda$ is the only closed additive subgroup of positive Lebesgue measure, therefore the Lebesgue measure is the only absolutely continuous stationary point of $I$.

\noindent
\emph{Proof of Theorem~\ref{stationary}.} 1. Let $\mu$ be the normalized Haar measure on an additive subgroup $G$, extended to $\Lambda$ with $\mu(\Lambda\setminus G)=0$. For arbitrary Borel sets $D_1\subset G$ and $D_2\subset\Lambda\setminus G$, and any $\x\in G$,
$$\mu(D_1-\x)=\mu(D_1),\quad \mu(D_2-\x)=\mu(D_2)=0.$$
Hence, for any Borel set $D\subset\Lambda$ and any $\x\in G$, $\mu(D-\x)=\mu(D)$ or, equivalently,
\be
(1_{D}*\mu)(\x)=\mu(D) \quad\mbox{for any $\x\in G$}.
\ee
Because any continuous function is the limit of a sequence of finite linear combinations of characteristic functions, (\ref{statf}) holds for every continuous $f$; hence, $\mu\in\calmst$.

2. Suppose that $\mu\in\calmst$ and let $A=\supp\mu$. Although the characteristic functions are discontinuous, Eq.~(\ref{statf}) extends to them through limits. For example, the characteristic functions of open sets are limits of monotone increasing sequences of continuous functions. For an arbitrary Borel set $D\subset\Lambda$,
\be
\mu(D+\x)\equiv\lambda_D\quad\mbox{for any $\x\in A$}
\ee
is then obtained by using the upper regularity of $\mu$.

(i) Suppose that $\0\in A$. Then $\mu(D+\x)=\mu(D)$ for all $D\subset\Lambda$ Borel sets and all $\x\in A$; in particular, for $D=A-\x$,
\be
\mu(A-\x)=\mu(A)=\mu(A\cap A-\x)=V
\ee
for all $\x\in A$. Since $A-\x$ is closed together with $A$, necessarily $A-\x=A$ for all $\x\in A$ and, hence, $A-A=A$. Thus, $A=\supp\mu$ is an additive subgroup of $\Lambda$ and $\mu$ is a shift-invariant measure on it.\\
(ii) If $\0\notin A$, choose any $\t\in -A$ and replace $A$ by $A+\t$ and $\mu$ by $T_\t\mu$. Because $T_\t\mu\in\calmst$, the argument (i) yields that $T_\t\mu$ is the normalized Haar measure on the additive group $A+\t$, and $\mu$ is a translate of this measure. $\Box$

\begin{corollary} Let $\mu,\nu\in\calm$ be two measures with coinciding supports. If both are stationary points then $\mu=\nu$.
\end{corollary}
\emph{Proof.} $\supp\mu=\supp\nu$ is an additive group and $\mu$ and $\nu$ are normalized Haar measures on it. But the normalized Haar measure is unique, whence $\mu=\nu$. $\Box$

\noindent
\emph{Remarks.} (i) Here are two examples showing that $\calmst\subsetneqq\calmst(u_\Lambda)$.\\
1. A two-dimensional example is the Dirac comb $\mu_H$ concentrated on the honeycomb lattice. It is not a stationary point because, if $\x_1$ and $\x_2$ are nearest neighbors, $\supp\mu_H-\x_2=-\supp\mu_H+\x_1$. However, $u_\Lambda(-\x)=u_\Lambda(\x)$ implies $\mu_H\in\calmst(u_\Lambda)$.\\
2. If $u$ is radial and $\Lambda=[-L/2,L/2]^d$, $u_\Lambda$ has a cubic symmetry. Then for any $0<a\leq L/4$
\be
\mu=\frac{V}{2d}\sum_{j=1}^d\left(\delta_{-a\e_j}+\delta_{a\e_j}\right)
\ee
is stationary for $u_\Lambda$, but not for functions not having a cubic symmetry.

(ii) If the periodic boundary condition is replaced by a free one, the set of stationary points changes completely. The reason is that now $u$ itself, and not the periodized $u_\Lambda$, appears in the energy functional; in fact, $\Lambda$ may not be a parallelepiped. The additive groups are replaced by others. For a radial $u$ the stationary points of $I_u$ are uniform distributions on 1, 2,..., d-1 dimensional spheres, on the vertices of regular polyhedra, in 3 dimensions also on the 60 vertices of the soccer ball (fullerene), and others. The supports are now sets that are invariant under transformations forming finite point groups or groups of rotation, and the stationary points correspond to Haar measures on those groups. As $\Lambda$ increases, the dependence of the ground states on the boundary condition should continuously disappear in the bulk. For example, on parallelepipeds there should be a continuous transition between IDGS's for free and periodic boundary conditions. Such a transition cannot take place on stationary points for both types of boundary conditions. Intuition suggests that stationary points have better chance with periodic boundary conditions.

(iii) If $u$ is integrable but $u(\x)\to\infty$ as $\x\to \0$ then $I[\mu]=\infty$ for any $\mu$ having a point part. However, depending on the rate of divergence of $u(\x)$, $I[\mu]$ can be finite in some continuous stationary points, and these are to be considered as possible IDGS's. If $u(\x)\sim |\x|^{-d+\zeta}$ near the origin with $0<\zeta<d$ then $u$ is integrable in $d-n$ dimensions for $n<\zeta$, so the energy functional is finite on the corresponding measures (\ref{lattice}). Some interactions not allowing stationary points of $I$ to be IDGS's will be presented in Theorem~\ref{theor-cusp}. If $0<\zeta<1$, $I[\mu]$ can be finite only if the Hausdorff dimension of $\supp\mu$ is greater than $d-\zeta>d-1$. According to our characterization given in Lemma~\ref{lem-Haar}, the only stationary point providing a finite energy for such an interaction is the Lebesgue measure. In Section~\ref{subsec-vgeq0} we will show, however, that the Lebesgue measure does not minimize $I$ if the Fourier transform of $u$ is partly negative. Thus, for $u\in L^1(\Rr^d)$ diverging at zero as $|\x|^{-d+\zeta}$ with $0<\zeta<1$ and having a partly negative Fourier transform, no stationary point is an IDGS. Theorem~\ref{theor-cusp} will cover this case.

\newsec{Ground state configurations in infinite space}\label{sec-GSC}

From Refs.~\cite{Su1,Su2} let us recall the definition of a ground state configuration in $\Rr^d$. To begin with, a configuration is an equivalence class of point sequences, two sequences being equivalent if they differ only in a permutation of the entries. We must consider sequences instead of sets because particles can be placed in the same point if the interaction is bounded. For an infinite configuration $X$ and a finite configuration $R$ let
\be\label{Ucond}
U(R|X)=U(R)+I(R,X)
\ee
with
\be
U(R)=\sum_{(\r,\r')\subset R}u(\r-\r'),\quad I(R,X)=\sum_{\r\in R,\x\in X}u(\r-\x).
\ee

\begin{definition} Let $u$ be a strongly tempered interaction. Given a real number $\mu$, an infinite configuration $X$ is a (grand canonical) ground state configuration for chemical potential $\mu$ (a $\mu$GSC) if for any finite configuration $X_f\subset X$, $U(X_f)$ is finite, the sum $I(X_f,X\setminus X_f)$ is absolutely convergent, and for any finite configuration $R\subset\Rr^d$,
\be\label{muGSC}
U(R|X\setminus X_f)-\mu|R|\geq U(X_f|X\setminus X_f)-\mu|X_f|.
\ee
$X$ is a (canonical) ground state configuration (GSC) if (\ref{muGSC}) holds true for every $R$ such that $|R|=|X_f|$.
\end{definition}
Thus, a ground state configuration is a locally stable configuration, in the sense that no local modification can decrease its energy. Equation~(\ref{muGSC}) must hold also when $R$ or $X_f$ is the empty set. Hence, if $X$ is a $\mu$GSC then
$$U(X_f|X\setminus X_f)-\mu|X_f|\leq 0$$
for any finite $X_f\subset X$. If $u$ is not strongly tempered, absolute convergence of $I(X_f,X\setminus X_f)$ must be replaced by some weaker condition, e.g. Abel summability, see \cite{Su2}. The above definition and the two lemmas below apply also to interactions that are not integrable at the origin or have a hard core.

The existence of GSC's and $\mu$GSC's is by no means obvious. For certain interactions they can be constructed explicitly ~\cite{Su1,Su2,Su4,Ven,Theil}, for some others only their existence is proven~\cite{Radin,BRS}. In this paper we assume that for the class of interactions considered, high-density GSC's do exist; at least, all statements about GSC's in infinite space are subject to this proviso.

\begin{definition} For $\x\in\Rr^d$, let $C_\x$ denote the unit ball centered at $\x$. An infinite configuration $X$ is locally uniformly finite (l.u.f.), if there exists an integer $m$ such that for any $\x\in\Rr^d$,
\be
|X\cap C_\x|\leq m.
\ee
Equivalently, for any compact $K\in\Rr^d$ there is an integer $m_K$ such that
\be
|X\cap(K+\x)|\leq m_K,\quad\mbox{any $\x\in\Rr^d$}.
\ee
\end{definition}
In \cite{Su2} we proved the absence of metastability in the case of strongly tempered interactions, by showing that a necessary condition for a l.u.f. configuration to be a GSC is to minimize the energy density among l.u.f. configurations of the same density. Here we prove that this condition is also sufficient among periodic arrangements (while it is obviously insufficient among l.u.f. configurations).

The general form of a periodic configuration is
\be\label{Xgen}
X=\biguplus_{j=1}^J(B+\y_j),
\ee
where $B=\Zz\a_1+\cdots+\Zz\a_d$, $\a_i\in\Rr^d$ are linearly independent, $\y_j$ are not necessarily different $d$-dimensional vectors, and in this paper we use $\biguplus$ to indicate that in the union coinciding points occur with repetition. The choice of $B$ can be made unique by asking $J$ to be minimal. $B$ thus obtained is the maximal group of period vectors of $X$, and is denoted by $B(X)$.

The particle density and the energy per particle of an infinite configuration $X$ are defined, respectively, as
\be\label{rho-ep}
\rho(X)=\lim_{\Omega\to\Rr^d}\frac{|X\cap\Omega|}{V(\Omega)},
\quad
e_p(X)=\lim_{\Omega\to\Rr^d}\frac{U(X\cap\Omega)}{|X\cap\Omega|},
\ee
provided that the limits exist. Here $\Omega$ is a bounded Lebesgue-measurable domain and $V(\Omega)$ is its volume. The energy density (energy per volume) of $X$ is $e(X)=\rho(X)e_p(X)$. If the particle density is kept fixed, we can work with any of $e_p$ or $e$. If $X$ is a periodic configuration, then $\rho(X)=|X\cap\Lambda|/V(\Lambda)$ where $\Lambda$ is any period-parallelepiped. If, moreover, $u$ is strongly tempered, cf. Eq.~(\ref{str}), and $\Omega$ tends to $\Rr^d$ in the Van Hove sense \cite{Rue}, then
\bea\label{epix}
e_p(X)=\frac{U(X_0)}{|X_0|}+\frac{1}{2|X_0|}\sum_{m=1}^\infty I(X_0,X_m)
=\frac{U(X_0)}{|X_0|}+\frac{I(X_0,X\setminus X_0)}{2|X_0|} .
\eea
Here $X_m (m=0,1,\ldots)$ are non-overlapping translates of any finite $X_0\subset X$ such that $X=\cup_{m=0}^\infty X_m$. The simplest choice is $X_0=(\y_1,\ldots,\y_J)$, cf. (\ref{Xgen}). However, $X_0$ can be the content of an arbitrarily large period cell, and then comparison of (\ref{epix}) and (\ref{rho-ep}) shows that
\be\label{I-small-X0}
I(X_0,X\setminus X_0)=o(|X_0|).
\ee
Since $u$ is superstable, for some $b\geq 0$ and $\rho(X)>b/C[u]$ we have
\be
|X_0|\leq \frac{U(X_0)}{-b+C[u]\rho(X)}
\ee
and, consequently,
\be\label{I-small-order}
\lim_{|X_0|\to\infty}\frac{I(X_0,X\setminus X_0)}{U(X_0)}=0.
\ee
The first part of the following lemma was proven by Sinai for a lattice gas~\cite{Sin} and by Radin in one dimensional continuous space~\cite{Rad84}.

\begin{lemma}\label{lemma-periodic-GSC}
Let the interaction be strongly tempered, and let $X$ be a periodic configuration. (i) If $X$ minimizes the energy density among periodic configurations of the same density and of period vectors belonging to $B(X)$, then $X$ is a GSC. (ii) If $X$ is a GSC then it minimizes the energy density among l.u.f. configurations of the same density (provided that the respective densities exist).
\end{lemma}
\emph{Proof.}
(i) Suppose that $X$ is not a GSC; then there exist $X_f\subset X$ and $R\subset\Rr^d$ finite sequences such that $R\cap X=\emptyset$, $|R|=|X_f|$, and
\be
\Delta=U(R|X\setminus X_f)-U(X_f|X\setminus X_f)<0.
\ee
We construct a periodic $Y$ whose period vectors form a subgroup of those of $X$, such that $\rho(Y)=\rho(X)$ and $e_p(Y)<e_p(X)$. For an odd integer $n$, let
\be
X_0=X\cap\Lambda_n,\qquad \Lambda_n = \left\{\sum_{i=1}^dx_i\a_i
:-\frac{n-1}{2}\leq x_i<\frac{n-1}{2}\right\}.
\ee
This is the union of the content of $n^d$ unit cells of $X$. Define $Y_0=X_0\setminus X_f\cup R$. Choose $n$ so large that $X_f\cup R\subset\Lambda_n$. Let $Y$ be the periodic extension of $Y_0$,
\be
Y=\bigcup_{\t\in n B}(Y_0+\t)=\biguplus_{\y\in Y_0} (n B+\y),
\ee
to be compared with
\be
X=\bigcup_{\t\in n B}(X_0+\t)=\biguplus_{\x\in X_0} (n B+\x).
\ee
It is straightforward to verify that
\be
\Delta=U(Y_0|X\setminus X_0)-U(X_0|X\setminus X_0).
\ee
Clearly, $\rho(X)=\rho(Y)$ and from (\ref{epix}) we find
\be
|X_0|[e_p(Y)-e_p(X)]=\Delta+\frac{1}{2}\left[I(Y_0,Y\setminus Y_0)-I(Y_0,X\setminus X_0)\right] + \frac{1}{2}\left[I(X_0,X\setminus X_0)-I(Y_0,X\setminus X_0)\right].
\ee
Now $X_0\setminus X_f=Y_0\setminus R$ implies
\be
I(X_0,X\setminus X_0)-I(Y_0,X\setminus X_0)=I(X_f,X\setminus X_0)-I(R,X\setminus X_0),
\ee
and, using $I(A,B+\t)=I(B,A-\t)$,
\be
I(Y_0,Y\setminus Y_0)-I(Y_0,X\setminus X_0)=I(R,Y\setminus Y_0)-I(X_f,Y\setminus Y_0).
\ee
The four infinite sums in the right member of the last two equations are absolutely convergent and tend to zero as $n$ tends to infinity, because the distances of $R$ and $X_f$ to $X\setminus X_0$ and to $Y\setminus Y_0$ diverge with $n$. Thus, if $n$ is large enough, $e_p(Y)<e_p(X)$.

\noindent
(ii) Earlier it was shown (Proposition, Ref.~\cite{Su2}) that if $X$ and $Y$ are two l.u.f. configurations of equal density and $e_p(X)>e_p(Y)$, then $X$ is not a GSC. Since periodic configurations are l.u.f., the second part of the claim follows. $\Box$

For a general parallelepiped $\Lambda$ spanned by the vectors $\l_i(i=1,\ldots,d)$, let
\be\label{def-ulam-gen}
u_\Lambda(\r)=\sum_{\t\in B_\Lambda}u\left(\r+\t\right)
\ee
where
\be\label{BLambda}
B_\Lambda=\left\{\sum_{i=1}^d n_i\l_i: n_1,\ldots,n_d\in\Zz\right\}.
\ee

\begin{lemma}\label{XcapL}
Let $u$ be strongly tempered and let $X$ be a periodic configuration. (i) If $X$ is a GSC of $u$, then $X\cap\Lambda$ is a ground state of $u_\Lambda$ on any period parallelepiped $\Lambda$ of $X$. (ii) If $X\cap\Lambda$ is a ground state of $u_\Lambda$ on any large enough period parallelepiped $\Lambda$ of $X$, then $X$ is a GSC of $u$.
\end{lemma}
\emph{Proof.} (i) If $X$ is a GSC, then it minimizes $e_p$ among periodic configurations of density $\rho(X)$. In particular, if $R\subset\Lambda$, $|R|=|X\cap\Lambda|$ and $Y$ is the periodic extension of $R$ from $\Lambda$, then $e_p(X)\leq e_p(Y)$. Comparison with Eq.~(\ref{epix}) written for a periodic configuration $Z$ of period cell $\Lambda$ and $X_0=Z\cap\Lambda$ shows that
\be\label{eLam/p}
e^\Lambda_p(Z\cap\Lambda):=\frac{U_\Lambda(Z\cap\Lambda)}{|Z\cap\Lambda|} =e_p(Z)-\frac{1}{2}\sum_{\0\neq\r\in B_\Lambda}u(\r).
\ee
Applying (\ref{eLam/p}) with $Z=X$ and $Y$, one finds
\be\label{gr-to-gr}
e^\Lambda_p(R)-e^\Lambda_p(X\cap\Lambda)=e_p(Y)-e_p(X)\geq 0.
\ee
(ii) We prove that if $Y$ is a periodic configuration, $B(Y)\subset B(X)$ and $\rho(Y)=\rho(X)$, then $e_p(Y)\geq e_p(X)$. Let $\Lambda$ be a period cell of $Y$ (and, hence, of $X$) chosen so large that $X\cap\Lambda$ is a ground state of $u_\Lambda$. Then inequality (\ref{gr-to-gr}) holds true with $R=Y\cap\Lambda$. $\Box$

\newsec{Infinite-density ground state in infinite space}\label{sec-IDGS-infinite}

Let ${\cal M}$ denote the family of (positive) Borel measures on $\Rr^d$.

\begin{definition}
Let $X_n$ be a sequence of infinite configurations with existing density $\rho_n$ and energy per particle $e_p(X_n)$ which tend to infinity with $n$. Suppose that
\be\label{ep/rho->Cu}
\lim_{n\to\infty}e_p(X_n)/\rho_n=C[u]
\ee
and there is some nonzero $\mu\in{\cal M}$ such that
\be\label{Dirac-comb}
\mu_{X_n}=\rho_n^{-1}\sum_{\x\in X_n}\delta_\x \rightharpoonup\mu.
\ee
Then $\mu$ is called an infinite-density ground state in $\Rr^d$.
\end{definition}
In contrast with IDGS's in finite volumes, the existence of IDGS's in infinite space is not \emph{a priori} guearanteed. The ultimate goal would be to prove the following commutative diagram:

\begin{conjecture}\label{conj-comdiag}
\begin{eqnarray}
\begin{CD}
    \mbox{GSC in $\Lambda$}@>\Lambda\to\infty>>\mbox{GSC in $\Rr^d$}\\
    @V\rho\to\infty VV@VV\rho\to\infty V\\
    \mbox{IDGS in $\Lambda$}@>>\Lambda\to\infty>\mbox{IDGS in $\Rr^d$}
\end{CD}
\end{eqnarray}
If $u$ is superstable and strongly tempered then high-density GSC's and IDGS's in infinite space exist. Any IDGS in $\Rr^d$ can be obtained as the vague limit of both a sequence of periodic extensions of IDGS's in increasing volumes and a sequence of Dirac combs (\ref{Dirac-comb}) associated with GSC's of increasing density. For any high-density GSC $X$, the associated measure $\mu_X$ is the vague limit of a sequence $\mu_{Y_n}$, where $Y_n$ is the periodic extension of a ground state of $u_{\Lambda_n}$ in some parallelepiped $\Lambda_n$, and $\rho(Y_n)=\rho(X)$.
\end{conjecture}
If the conjecture holds true then information on GSC's in $\Rr^d$ can be obtained by studying IDGS's in finite volume. In the case when $v\geq 0$ and of compact support the results of Refs.~\cite{Su1,Su2} and those of Section~\ref{subsec-vgeq0} below prove the conjecture but nothing new can be obtained about GSC's. If $v\geq 0$ but its support is noncompact, the existence of IDGS's in infinite space still follows from $C_\Lambda[u]\equiv C[u]=v(\0)/2$; the Lebesgue measure is one of the IDGS's (the only one if $v>0$) and Section~\ref{subsec-vgeq0} gives a fairly complete description of them, cf. Corollary~\ref{cor-IDGS-inf}. Concerning the general case, below we present some clarification and partial result. Let $\Lambda$ be any parallelepiped centered at zero,
\be
\Lambda=\left\{\sum_{i=1}^dx_i\l_i: -1/2\leq x_i<1/2, i=1,\ldots,d\right\}
\ee
where $\l_i$ are linearly independent vectors. If $n$ is a positive integer, then $n\Lambda=\{n\x:\x\in\Lambda\}$ is the union of $n^d$ translates of $\Lambda$:
\be
n\Lambda=\bigcup_{j=1}^{n^d}(\Lambda+\t_j)
\ee
where
\be\label{tj}
\t_j=\frac{1}{2}\sum_{i=1}^d m_{ji}\l_i,\quad m_{j1},m_{j2},\dots,m_{jd}\in\{-n+1,-n+3,\ldots,n-3,n-1\}.
\ee

\begin{definition}
Let $n$ be a positive integer. The periodic extension $E_n$ from  $\calm$ to ${\cal M}_{n\Lambda}$ is a map assigning
\be\label{Ext-n}
E_n\mu=\sum_{j=1}^{n^d}1_{\Lambda+\t_j}T_{\t_j}\mu
\ee
to $\mu\in\calm$; that, is,
\be
E_n\mu(D)=\mu(D-\t_j),\quad j=1,\ldots,n^d
\ee
for Borel sets $D\subset\Lambda+\t_j$. The periodic extension $E$ from $\calm$ to ${\cal M}$ is a map assigning
\be\label{Ext}
E\mu=\sum_{\t\in B_\Lambda}1_{\Lambda+\t}T_\t\mu
\ee
to $\mu\in\calm$.
\end{definition}

\begin{lemma}
$E\mu$ is the vague limit of $E_{2n+1}\mu$ as $n$ tends to infinity.
\end{lemma}
\emph{Proof.} Let $f\in C_c(\Rr^d)$. Comparing (\ref{Ext}) with (\ref{Ext-n}) one obtains
\be
E\mu(f)=\lim_{n\to\infty}E_{2n+1}\mu(f).\quad \Box
\ee

\noindent
Recall that $\Lambda^*=\{\sum n_i\l_i':n_i\in\Zz,i=1,\ldots,d\}$, where $\l_i'\cdot\l_j=2\pi\delta_{ij}$. Clearly,
\be\label{lam-in-lam}
\Lambda^*\subset(n\Lambda)^*=n^{-1}\Lambda^*.
\ee

\begin{lemma}\label{lem-perext}
If $n$ is a positive integer and $\k\in\Lambda^*$ then
\be
\widehat{E_n\mu}(\k)=\left\{\begin{array}{cl}
                     \widehat{\mu}(\k)& \mbox{if $n$ is odd}\\
                     s(\k)\widehat{\mu}(\k)&\mbox{if $n$ is even}
                           \end{array}\right.
\ee
where $s(\k)\in\{-1,1\}$. If $\k\in (n\Lambda)^*\setminus\Lambda^*$ then $\widehat{E_n\mu}(\k)=0$. Furthermore,
\be
\widehat{E\mu}(\k)=1_{\Lambda^*}(\k)\widehat{\mu}(\k),\quad\k\in\Rr^d.
\ee
\end{lemma}
\emph{Proof.} Using the definition of $E_n\mu$ and (\ref{muhat}),
\be
\widehat{E_n\mu}(\k)=\frac{1}{n^dV(\Lambda)} \int_{n\Lambda}e^{-i\k\cdot\x}E_n\mu(\d\x) =\frac{1}{n^dV(\Lambda)}\sum_{j=1}^{n^d} \int_{\Lambda}e^{-i\k\cdot(\x+\t_j)}\mu(\d\x)
=\frac{1}{n^d}\sum_{j=1}^{n^d}e^{-i\k\cdot\t_j}\widehat{\mu}(\k).
\ee
The average of the complex units is 1, $\pm 1$, or 0, depending on the case. Taking the limit $n\to\infty$ gives the result for $\widehat{E\mu}(\k)$. $\Box$

Below we use the notations $I_\Lambda[\mu]$ and $I^0_\Lambda$ for the energy functional (\ref{Imu}) and its infimum (\ref{I0}), and $\epsilon_\Lambda(\r)_N$, $\epsilon_{\Lambda,N}$, $\omega_\Lambda(\r)_N$ and $\omega_{\Lambda,N}$ for (\ref{omega'rn}), (\ref{omega'n}), (\ref{omegarn}) and (\ref{omegan}), respectively, to indicate the dependence on $\Lambda$. Expanding the sum that defines $u_\Lambda$,
\be\label{I-extended}
I_\Lambda[\mu]=\frac{1}{2}\int_{\Rr^d}u(\r)\d E\gamma_\Lambda(\r)=\frac{1}{2}E\gamma_\Lambda(u)
\ee
where $\gamma_\Lambda=\mu*\widetilde{\mu}/V$ for $\mu\in\calm$.

\begin{proposition}
For any positive integer $n$, $C_{n^m\Lambda}[u](m=1,2,\ldots)$ is a monotone decreasing sequence.
\end{proposition}
\emph{Proof.} Let $\Lambda_m= n^m\Lambda$. From (\ref{I-extended}) or from Lemma~\ref{lem-perext} and the form (\ref{nonneg}) of the energy functional, it follows that for any $\mu\in{\cal M}_{\Lambda_m}$
\be
I_{\Lambda_{m+1}}[E_n\mu]=I_{\Lambda_m}[\mu].
\ee
By Theorems~\ref{bounded} and \ref{unbounded} then
\be
C_{\Lambda_{m+1}}[u]=I^0_{\Lambda_{m+1}}
\leq I^0_{\Lambda_{m}}=C_{\Lambda_{m}}[u].\quad \Box
\ee

For a stable interaction $C_\Lambda[u]\geq 0$ for all $\Lambda$, therefore the limit of $C_{n^m\Lambda}[u]$ as $m$ tends to infinity exists and is bounded below by the best superstability constant $C[u]$. As noted earlier, the limit must, in fact, be $C[u]$. Now we prove a different form of $C[u]$.

\begin{proposition}\label{prop-C[u]} If $u$ is bounded then
\be\label{Cu-ep/rho}
C[u]=\lim_{\rho\to\infty}\frac{1}{\rho}\inf\left\{e_p(X):\mbox{$X$ periodic, $\rho(X)=\rho$}\right\}.
\ee
\end{proposition}

\noindent
\emph{Remark.} Here we have not supposed that there are periodic GSCs for arbitrarily high densities. If this is the case, the infimum in (\ref{Cu-ep/rho}) is attained, and $C[u]=\lim_{n\to\infty}e_p(X_n)/\rho(X_n)$ where $X_n$ is any sequence of periodic GSCs with $\rho(X_n)$ tending to infinity.

\noindent
\emph{Proof.} (i) First we find a sequence $X_n$ of periodic configurations such that $\rho(X_n)$ diverges and $e_p(X_n)/\rho(X_n)$ tends to $C[u]$. Let $\Lambda_n$ be any sequence of parallelepipeds such that $C_{\Lambda_n}[u]$ converges to $C[u]$. For each $n$ let $R_{nm}(m=1,2,\ldots)$ be a sequence of ground state configurations in $\Lambda_n$, $|R_{nm}|$ going to infinity with $m$. According to Equation~(\ref{epsilon-limit}), $\lim_{m\to\infty}\epsilon_{\Lambda_n}(R_{nm})=C_{\Lambda_n}[u]$. Therefore, one can choose a subsequence $R_{nm_n}\equiv R_n$ such that $\lim\epsilon_{\Lambda_n}(R_n)=\lim\omega_{\Lambda_n}(R_n)=C[u]$. Let $X_n$ denote the periodic extension of $R_n$. Thus, $R_n=X_n\cap\Lambda_n$ and $|R_n|/V(\Lambda_n)=\rho(X_n)\to\infty$. From Eq.~(\ref{eLam/p}), for any periodic configuration $X$ of period parallelepiped $\Lambda$
\be\label{epix-eps}
\frac{e_p(X)}{\rho(X)}=\frac{|X\cap\Lambda|-1}{|X\cap\Lambda|}\,\epsilon_{\Lambda}(X\cap\Lambda)+ \frac{1}{\rho(X)}\sum_{\0\neq\r\in B_{\Lambda}}u(\r)=\omega_\Lambda(X\cap\Lambda)-\frac{u(\0)}{2\rho(X)}.
\ee
Applying this formula to $X_n$ and $\Lambda_n$, we obtain $\lim_{n\to\infty}e_p(X_n)/\rho(X_n)=C[u].$\\
(ii) Suppose there is a sequence $Y_n$ of periodic configurations with respective period cells $\Lambda_n$ such that $\rho(Y_n)$ tends to infinity and
$$\lim_{n\to\infty} \frac{e_p(Y_n)}{\rho(Y_n)}=C_0< C[u].$$
We can always choose the sequence $\Lambda_n$ so that it tends to infinity in Fisher's sense. We may also suppose that $Y_n\cap\Lambda_n$ is a ground state of $u_{\Lambda_n}$, otherwise we can replace $Y_n$ by the periodic extension of an $|Y_n\cap\Lambda_n|$-particle ground state of $u_{\Lambda_n}$, cf. Eq.~(\ref{gr-to-gr}). For any $\varepsilon>0$ there is a $N$ such that for any $n\geq N$,
$$
\left|\frac{e_p(Y_n)}{\rho(Y_n)}-C_0\right|\leq\frac{\varepsilon}{3},\quad \left|\epsilon_{\Lambda_n}(Y_n\cap\Lambda)-\frac{e_p(Y_n)}{\rho(Y_n)}\right|\leq\frac{\varepsilon}{3}, \quad C_{\Lambda_n}[u]-\epsilon_{\Lambda_n,|Y_n\cap\Lambda|}\leq\frac{\varepsilon}{3}.
$$
The second and third inequalities use Eq.~(\ref{epix-eps}) and Corollary~\ref{coro-uniform}, respectively. By assumption,
$$\epsilon_{\Lambda_n}(Y_n\cap\Lambda)=\epsilon_{\Lambda_n,|Y_n\cap\Lambda|},$$
therefore
$$|C_{\Lambda_n}[u]-C_0|\leq\varepsilon\qquad\mbox{if $n\geq N$}.$$
Thus, $C_{\Lambda_n}[u]$ converges to $C_0<C[u]$, but this contradicts the hypothesis (\ref{Cu-Fisher}). $\Box$

\noindent
\emph{Remarks.} (i) Without the hypothesis (\ref{Cu-Fisher}), from (\ref{Cu}) we still find that equation~(\ref{Cu-ep/rho}) holds true if $\lim_{\rho\to\infty}$ is replaced by $\liminf_{\rho\to\infty}$. (ii) The first equality of Equation~(\ref{epix-eps}) applies also to unbounded interactions. Equation~(\ref{Cu-ep/rho}) is presumably valid in this case as well, but in the absence of a proof of Corollary~\ref{coro-uniform} for unbounded interactions we cannot prove it.

For the sequence $X_n$ used in the proof of Proposition~\ref{prop-C[u]}, Eq.~(\ref{ep/rho->Cu}) holds true. However, the proof of Eq.~(\ref{Dirac-comb}) is missing. We may suppose that $\Lambda_n=(2l_n+1)Q_\0$, where $Q_\0$ is the unit cube centered at $\0$ and the integers $l_n$ tend to infinity with $n$. $X_n$ can be selected so that $|X_n\cap Q_\0|=\max_{\t\in\Rr^d}|(X_n+\t)\cap Q_\0|$, then $\mu_{X_n}(Q_\0)\geq V(Q_\0)=1$ because $\mu_{X_n}(\Lambda_n)= V(\Lambda_n)$. The task would be to prove that $\mu_{X_n}(Q_\0)$ is a bounded sequence.  If this holds, one could find a vaguely convergent subsequence of $\mu_{X_n}$ tending to a nonzero measure which is by definition an IDGS in $\Rr^d$. The same IDGS in $\Rr^d$ could also be obtained as the vague limit of a sequence of IDGS's $\mu_n\in{\cal M}_{\Lambda_n}$ chosen such that $\mu_n(Q_\0)\geq V(Q_\0)$.

Let $G$ be a locally compact Abelian group (in this paper $G$ is a torus or $G=\Rr^d$). A measure $\mu$ on $G$ is called shift-bounded if for any compact $A\subset G$ there is a constant $\alpha_A$ such that for all $\a\in G$,
\be\label{shift-bounded}
\mu(A+\a)\leq\alpha_A.
\ee
Suppose again that $\Lambda_n$ is a sequence of increasing parallelepipeds such that $C_{\Lambda_n}[u]$ converges to $C[u]$. For any $n$ let $\mu_{n}$ be an IDGS in $\Lambda_n$. Hence, with the notation $\gamma_{n}=V(\Lambda_n)^{-1}\mu_{n}*\widetilde{\mu_{n}}$,
\be\label{gamma-n}
\frac{1}{2}\lim_{n\to\infty}\gamma_{n}(u_{\Lambda_n})= \frac{1}{2}\lim_{n\to\infty}E\gamma_{n}(u) = \frac{1}{2}\lim_{n\to\infty}\widehat{E\gamma_{n}}(v) = \frac{1}{2}\lim_{n\to\infty}\widehat{\gamma_{n}}(v)=C[u].
\ee
$E\gamma_n$ is periodic and, thus, shift-bounded. $\widehat{E\gamma_n}$ is a positive measure which is the Fourier transform of a measure, therefore it is also shift-bounded, see \cite{BF} (Proposition~4.9) or \cite{Hof} (Proposition~3.3). A necessary condition for (\ref{Dirac-comb}) to hold is that these sequences are uniformly shift-bounded. In a special case the proof is given in the following lemma.

\begin{lemma}
If $u(\x)\geq c e^{-\alpha|\x|^2}$ for some $c,\alpha>0$, then both sequences $E\gamma_n$ and $\widehat{E\gamma_{n}}$ are uniformly shift-bounded.
\end{lemma}
\emph{Proof.}
Let $A$ be a unit cube. If $g(\k)=Ke^{-|\k|^2/4\alpha}$ with $K=\left(\int_0^1e^{-q^2/4\alpha}\d q\right)^{-d}$ then
$1_A*g\geq 1_A$. $\breve{f}$ denoting the inverse Fourier transform of $f$,
\bea\label{shift-bounded1}
\sum_{\k\in\Lambda_n^*\cap A+\a}|\widehat{\mu_n}(\k)|^2= \widehat{E\gamma_{n}}(A+\a)\leq\widehat{E\gamma_{n}}(1_{A+\a}*g)= E\gamma_n(\breve{1}_{A+\a}\breve{g})\leq E\gamma_n(|\breve{1}_{A+\a}|\breve{g})\nonumber\\
\leq \frac{K}{(2\pi)^d(\pi/\alpha)^{d/2}} \int e^{-\alpha|\x|^2}\d E\gamma_n(\x)\leq\beta\int u(\x)\d E\gamma_n(\x)\leq \beta C[u]
\eea
for $n$ sufficiently large ($\beta=\frac{K}{c(2\pi)^d(\pi/\alpha)^{d/2}}$). The passage to any compact $A$ is obvious, so $\widehat{E\gamma_{n}}$ is uniformly shift-bounded. From (\ref{shift-bounded1}) or with a similar independent argument one can show that for any Schwartz function $f$, $\widehat{E\gamma_{n}}(f)(n=1,2,\ldots)$ is a bounded sequence. We use this fact to prove that $E\gamma_n$ is uniformly shift-bounded. Let again $A$ be a unit cube and $g(\x)$ as before; then there is some constant $c_g$ such that
\be
E\gamma_{n}(A+\a)\leq E\gamma_{n}(1_{A+\a}*g)=\widehat{E\gamma_n}(\widehat{1}_{A+\a}\widehat{g}) \leq \widehat{E\gamma_n}(|\widehat{1_{A}}|\widehat{g})\leq \widehat{E\gamma_n}(\widehat{g}) \leq c_g\quad\mbox{all $n$}.\quad\Box
\ee

An example to this lemma is the Gaussian pair potential for which, as mentioned already, the unique IDGS is the Lebesgue measure. Other examples are pair potentials with a Gaussian lower bound and a partly negative Fourier transform -- the latter condition holds, for example, if $\Delta u(\0)=0$, cf. Eq.~(\ref{cusp-0}). Unfortunately, uniform shift-boundedness of the sequences $E\gamma_n$ and $\widehat{E\gamma_{n}}$ does not allow one to conclude that the sequence $\mu_n$ is uniformly shift-bounded. Unbounded but $o(\sqrt{V(\Lambda_n)})$ weight rearrangements of $\mu_n$ do not affect uniform shift-boundedness of $E\gamma_n$.

\newsec{Infinite-density ground states for $v\geq 0$}\label{subsec-vgeq0}

The most complete information about IDGS's can be obtained when the Fourier transform $v$ of the interaction is nonnegative. The minimum of $I[\mu]$ is attained on any $\mu$ not contributing to the sum in Eq.~(\ref{nonneg}), and its value is $I^0=v(\0)/2$, thus, $C_\Lambda[u]=v(\0)/2=C[u]$ for every $\Lambda$.

\begin{proposition}\label{prop-vgeq0} The Lebesgue measure $\lambda$ is an IDGS in $\Lambda$ if and only if $v\geq 0$, and then it yields
\be\label{v>0}
I[\lambda]=C[u]=v(\0)/2.
\ee
If $v>0$, the Lebesgue measure is the unique IDGS.
\end{proposition}
\emph{Proof.} For $\k\in\Lambda^*$,
\be
\widehat{\lambda}(\k)=\delta_{\k,\0},
\ee
and $\lambda$ is the only element of $\calm$ with Fourier transform $\delta_{\k,\0}$. So $\lambda$ is an IDGS, and is the only one if $v$ is strictly positive. It remains to prove that the Lebesgue measure is not an IDGS if $v$ is partly negative. In this case consider
\be\label{muq}
\d\mu_\q(\x)=(1+\cos\q\cdot\x)\d\x
\ee
and choose $\q\in\Lambda^*$ such that $v(\q)=\min_{\k\in\Lambda^*}v(\k)<0$. Then
\be
\widehat{\mu_\q}(\k)=\delta_{\k,\0}+\frac{1}{2}\left(\delta_{\k,\q}+\delta_{\k,-\q}\right)
\ee
and, recalling that $v(-\k)=v(\k)$,
\be\label{Imuq}
I^0\leq I[\mu_\q]={v(\0)\over 2}+{v(\q)\over 4}<{v(\0)\over 2}=I[\lambda],
\ee
so $\lambda$ is not an IDGS. $\Box$

Let us make a small detour here to explain the choice of $\mu_\q$. The energy functional $I[\mu]$ is the diagonal part of a quadratic form on $\calm$. The operator underlying it has nice properties.

\begin{proposition} Let $u$ be an integrable even pair potential, $u(-\x)=u(\x)$. On $L^2(\Lambda)$ define an integral operator $A$ by setting
\be\label{opA}
(A\phi)(\x)=\int_\Lambda u_\Lambda(\x-\y)\phi(\y)\d\y.
\ee
Then $A$ is a bounded self-adjoint operator with eigenvalues $\{v(\k):\k\in\Lambda^*\}$ and eigenvectors $\psi_\k(\x)=\exp(i\k\cdot\x)$.
\end{proposition}
\emph{Proof.} If $\k\in\Lambda^*$ then
\be
v(\k)=\int u(\x)e^{-i\k\cdot\x}\d\x=\int_\Lambda u_\Lambda(\x)e^{-i\k\cdot\x}\d\x.
\ee
Substituting $\phi=\psi_\k$ into Eq.~(\ref{opA}) gives $A\psi_\k=v(\k)\psi_\k$. $\Box$

\noindent
Thus, $\mu_\q$ is an absolutely continuous measure composed of the lowest-lying real eigenvector of $A$ that we must mix with the constant eigenvector in order to make the sum nonnegative.

If $v\geq 0$ and takes on the zero value, besides $\lambda$ many other IDGS's appear. We consider two specific cases.

\begin{proposition}\label{prop-vgeq-piling}
Suppose that $v\geq 0$ and $v(\k)=0$ for $|\k|\geq K_0$, where $K_0<\infty$. Let $\Lambda$ be a parallelepiped, $B$ any Bravais lattice such that $\Lambda$ is a period cell for $B$, and let $q_{B^*}$ denote the nearest-neighbor distance of $B^*$, the reciprocal lattice of $B$. If $q_{B^*}\geq K_0$ then
\be\label{muB}
\mu_{B\cap\Lambda}=\frac{V}{|B\cap\Lambda|}\sum_{\x\in B\cap\Lambda}\delta_\x
\ee
is an IDGS.
\end{proposition}
\emph{Remarks.} (i) For the finite point configurations associated with $\mu_{B\cap\Lambda}$, the proposition partly repeats the results of \cite{Su1}. It provides also an example to Proposition~\ref{prop-piling}. (ii) $\mu_{B\cap\Lambda}$ extends periodically to $\Rr^d$ into
\be
\mu_B=\rho(B)^{-1}\sum_{\x\in B}\delta_\x,
\ee
and the energy functional shown below actually depends only on $\mu_B$.

\noindent
\emph{Proof.} For $\mu_{B\cap\Lambda}$ given by (\ref{muB}) and for $\k\in\Lambda^*$,
\be
\widehat\mu_{B\cap\Lambda}(\k)=
1_{B^*}(\k)=
\sum_{\K\in B^*}\delta_{\k,\K}.
\ee
This implies
\be\label{I[muB]}
I[\mu_{B\cap\Lambda}]={1\over 2}\left[v(\0)+\sum_{\0\neq\K\in B^*}v(\K)\right]\equiv I[\mu_B]
\ee
The shortest nonzero vector in $B^*$ has a length $q_{B^*}$, at and above which $v(\k)$ vanishes, therefore $I[\mu_{B\cap\Lambda}]=v(\0)/2=I^0$. $\Box$

The second special case is an interaction with a ``soft mode'', $v(\k)=0$ for $|\k|=k_0$ and $v$ is positive otherwise. One may think of the example
\be
v(\k)=c_1e^{-c_2|\k|}(|\k|-k_0)^2
\ee
with some positive constants $c_1,c_2,k_0$, but the precise form of $v$ plays no role. Choose $\Lambda$ so that $\Lambda^*$ contains a vector $\q$ of length $k_0$; then there are at least two vectors, $\pm\q$ of length $k_0$. The following proposition is an immediate consequence of the discussion above. The measure $\mu_\q$ appearing in it is an example for an IDGS that is not a stationary point of $I$.

\begin{proposition} If $v(\k)\geq 0$ and $v(\k)=0$ for $|\k|=k_0$, the measure $\mu_\q$ of Eq.~(\ref{muq}) with $|\q|=k_0$ is an IDGS.
\end{proposition}

\noindent
A particularity of the case $v\geq 0$ is that convex combinations of IDGS's are also IDGS's.

\begin{lemma} If $v\geq 0$ then $I[\mu]$ is a convex functional on $\calm$, that is, if $\mu,\nu\in\calm$ and $0< \alpha,\beta< 1$, $\alpha+\beta=1$ then
\be
I[\alpha\mu+\beta\nu]\leq\alpha I[\mu]+\beta I[\nu].
\ee
\end{lemma}
\emph{Proof.} This immediately follows from the form (\ref{nonneg}) of $I[\mu]$ and the convexity of $|z|^2$ on the complex plane:
\be
|\alpha z_1+\beta z_2|^2\leq(\alpha|z_1|+\beta|z_2|)^2\leq\alpha|z_1|^2 +\beta|z_2|^2.\qquad\Box
\ee

\begin{proposition} If $v\geq 0$, the set of IDGS's is convex.
\end{proposition}
\emph{Proof.} Let $\mu,\nu\in\calm$ be two IDGS's and $0< \alpha,\beta< 1$, $\alpha+\beta=1$.
\be
I^0\leq I[\alpha\mu+\beta\nu]\leq\alpha I[\mu]+\beta I[\nu]=v(\0)/2=I^0,
\ee
therefore $\alpha\mu+\beta\nu$ is an IDGS. $\Box$

\begin{corollary}\label{cor-IDGS-inf}
If $v\geq 0$ then the periodic extension of any IDGS from any parallelepiped $\Lambda$ is an IDGS in $\Rr^d$. IDGS's in $\Rr^d$ form a convex set. If $v>0$, the Lebesgue measure is the unique IDGS in $\Rr^d$.
\end{corollary}

\newsec{Uniformity vs nonuniformity of high-density ground states}\label{sec-unif-nonunif}

Proposition~\ref{prop-vgeq0} has an immediate implication on the distribution of particles in high-density ground states in $\Lambda$. If $\mu$ is an IDGS and $R_m$ is a sequence of $N_m$-particle ground states in $\Lambda$ such that $\mu_{R_m}$ weakly converges to $\mu$, then
\be\label{limD}
\lim_{m\to\infty}\frac{|R_m\cap D|}{\rho(R_m)}=\mu(D)
\ee
for any open $D\subset\Lambda$, where $\rho(R_m)=N_m/V$. Equation~(\ref{limD}) is a direct consequence of the definition of an IDGS. If $v>0$ then $\lambda$ is the unique IDGS. Thus, for any sequence $R_N$ of $N$-particle ground states $\mu_{R_N}$ converges weakly to the Lebesgue measure $\lambda$, resulting
\be
\lim_{N\to\infty}\frac{|R_N\cap D|}{\rho(R_N)}=\lambda(D).
\ee
On the other hand, if $v$ is partly negative, the asymptotic distribution of particles in any ground state is necessarily inhomogeneous. We shall prove an analogous result for ground state configurations in infinite space.

In Refs.~\cite{Su1,Su2} we gave a presumably complete description of defect-free high-density GSC's of interactions with a nonnegative Fourier transform of compact support. If the Fourier transform is strictly positive,  a prominent example of which is the Gaussian pair potential, a rigorous identification of GSC's is still missing and, if the decorrelation conjecture of Torquato and Stillinger~\cite{TS1} is valid, the task is near to impossible in high dimensions. One is then reduced to some probabilistic description, and such an approach may be helpful already for $d< 8$. According to another conjecture of Torquato and Stillinger~\cite{TS2}, for the Gaussian potential at $d\leq 8$ the close-packed Bravais lattice and its reciprocal lattice should be the unique GSC at low and high densities, respectively. However, this conjecture was disproved by Cohn and Kumar~\cite{CK}, who found uniform close-packed periodic structures of lower energy at low densities in 5 and 7 dimensions. The result below shows the fundamental difference between interactions with a strictly positive or a partly negative Fourier transform. It may also be helpful in the numerical search for low-energy arrangements at high densities.

\begin{theorem}\label{hom-inhom}
For arbitrary $d\geq 1$, let $u$ be a bounded, integrable, strongly tempered pair potential. Suppose that there exists a sequence of periodic GSC's $X_n$ of respective densities $\rho_n$ tending to infinity, and let
\be\label{muxn}
\mu_{X_n}=\rho_n^{-1}\sum_{\x\in X_n}\delta_\x.
\ee
Write $X_n$ in the form (\ref{Xgen}),
\be\label{Xn-periodic}
X_n=\biguplus_{j=1}^{J_n}(B_n+\y_j),
\ee
where $B_n=B(X_n)=\sum_{i=1}^d\Zz\a_{ni}$.

\noindent
(i) Let the Fourier transform $v$ of the potential be strictly positive. Suppose that the sequence of lattice constants $|\a_{ni}|(i=1,\ldots,d;n=1,2,\ldots)$ is bounded. Then $\mu_{X_n}$ converges to the Lebesgue measure in distribution sense, that is, for any Schwartz function $f:\Rr^d\to\Cc$,
\be\label{distr-limit}
\lim_{n\to\infty}\int f(\x)\d\mu_{X_n}(\x)=\int f(\x)\d\x.
\ee

\noindent
(ii) If $v$ is partly negative, then (\ref{distr-limit}) does not hold for any subsequence of $\mu_{X_n}$.
\end{theorem}
\emph{Remarks.} The restriction to bounded interactions is for technical reasons. The result implies that if $v$ is partly negative, any (vaguely or in distribution sense) convergent subsequence of $\mu_{X_n}$ tends to an IDGS in infinite space which is different from the Lebesgue measure. The IDGS's are expected to be periodic or almost periodic in this case. Accordingly, the distribution of particles in high-density GSC's shows a pattern corresponding to the infinite-density limit. On the other hand, if $v>0$ then the distribution of particles is asymptotically uniform. We cannot prove that the set of lattice constants is always bounded, but the opposite, the divergence of at least one of the $d$ lattice constants, leaves only two possibilities. The first is that $J_n$, the number of particles in the primitive cell (called complexity), tends to infinity. The second is that $J_n$ is bounded, implying that $X_n$ falls into the union of lower than $d$-dimensional substructures of diverging separation and density. In the case of the Gaussian potential, in 5 and 7 dimensions, a new numerical result~\cite{CKS} suggests a uniaxially anisotropic high-density GSC with a stronger compression along the distinguished axis than perpendicular to it. According to part (i) of the theorem, the other lattice constants cannot remain bounded if the anisotropy is to survive the limit of infinite density.

\noindent
\emph{Proof.} (i) Let $\Lambda_n$ be a period parallelepiped of $B_n$ and, hence, of $X_n$. Then, by Lemma~\ref{XcapL}, $X_n\cap\Lambda_n$ is a ground state of $u_{\Lambda_n}$.
From Eqs.~(\ref{om-eps}), (\ref{sup-C-inf}) and (\ref{v>0}),
\be
0<\omega_{|X_n\cap\Lambda_n|}-\frac{v(\0)}{2}<\frac{u_{\Lambda_n}(\0)}{2\rho_n}.
\ee
On the other hand,
\be
\omega_{|X_n\cap\Lambda_n|}=I[\mu_{X_n}]=\frac{v(\0)}{2}+\frac{1}{2} \sum_{\0\neq\k\in B_n^*}v(\k)|\widehat{\mu}_{X_n}(\k)|^2,
\ee
where
\be
\widehat{\mu}_{X_n}(\k)=J_n^{-1}\sum_{j=1}^{J_n}e^{i\k\cdot\y_j}\quad\mbox{if $\k\in B_n^*$.}
\ee
Thus,
\be
\sum_{\0\neq\k\in B_n^*}v(\k)|\widehat{\mu}_{X_n}(\k)|^2 <\frac{u_{\Lambda_n}(\0)}{\rho_n}<\frac{2u(\0)}{\rho_n}
\ee
for $n$ large enough; moreover, for any bounded set $D$
\be\label{D}
\sum_{\0\neq\k\in B_n^*\cap D}|\widehat{\mu}_{X_n}(\k)|^2 <\frac{2u(\0)}{\rho_n\inf_{\k\in D}v(\k)}.
\ee
Take any $f\in\mathcal{S}(\Rr^d)$ (function of rapid decrease). Using $\rho_n=J_n\rho(B_n)$, the Poisson summation formula yields
\be
\int f\d\mu_{X_n}=\rho_n^{-1}\sum_{\x\in X_n}f(\x)=\int f(\x)\d\x
+\sum_{\0\neq\k\in B_n^*}\widehat{f}(\k)\widehat{\mu}_{X_n}(\k).
\ee
Because the set of period lengths of $B_n^*(n=1,2,\ldots)$ is separated from zero, the sum over $\k\neq\0$ tends to zero as $n$ goes to infinity. Indeed, for $\m\in\Zz^d$, let
\be
Q_\m=\{\k\in\Rr^d: m_i\leq k_i\leq m_i+1\,(i=1,\ldots,d)\}\setminus\{\0\}.
\ee
Then $\cup_\m Q_\m=\Rr^d\setminus\{\0\}$ and $|B^*_n\cap Q_\m|\leq l$ with some $l$ independent of $n$ and $\m$. Since $v$ is continuous and strictly positive, for any $M>0$ there exists an $\varepsilon>0$ such that $v(\k)\geq\varepsilon$ if $\k\in\cup_{|\m|\leq M}Q_\m$. Write
\be
\sum_{\0\neq\k\in B_n^*}=\sum_{\k\in B^*_n\cap(\cup_{|\m|\leq M}Q_\m)} + \sum_{\k\in B^*_n\cap(\cup_{|\m|>M}Q_\m)}.
\ee
Using (\ref{D}), by Cauchy's inequality
\be
\left|\sum_{\k\in B^*_n\cap(\cup_{|\m|\leq M}Q_\m)}\widehat{f}(\k)\widehat{\mu}_{X_n}(\k)\right|^2
\leq \frac{2u(\0)}{\rho_n\varepsilon} \sum_{\k\in B^*_n\cap(\cup_{|\m|\leq M}Q_\m)}|\widehat{f}(\k)|^2\leq\frac{2u(\0)lM^d\|\widehat{f}\|_\infty^2}{\rho_n\varepsilon}
\ee
which tends to zero as $n\to\infty$. On the other hand,
\be
\left|\sum_{\k\in B^*_n\cap(\cup_{|\m|> M}Q_\m)}\widehat{f}(\k)\widehat{\mu}_{X_n}(\k)\right|\leq l\sum_{|\m|>M}\max_{\k\in Q_\m} |\widehat{f}(\k)|.
\ee
The sum on the right-hand side is convergent and tends to zero as $M$ tends to infinity, proving Eq.~(\ref{distr-limit}).

\noindent(ii)
Fix $Q=[-L/2,L/2]^d$ and define $N_n=\lfloor\rho_n L^d\rfloor$. Let $Q_n=[-L_n/2,L_n/2]^d$ where $L_n^d=N_n/\rho_n$, then $0\leq L^d-L_n^d<\rho_n^{-1}$. Let $\q\in Q^*=(2\pi/L)\Zz^d$ be the minimizer of $v(\k)$ in $Q^*$, and let $\q_n=(L/L_n)\q\in Q^*_n=(2\pi/L_n)\Zz^d$. Because $L_n\to L$, $\q_n$ tends to $\q$ as $n$ increases. Let $\mu_{R_n}\in{\cal M}_{Q_n}$ be a $N_n$-point approximation of $\mu_{\q_n}$, cf. Eq.~(\ref{muq}), constructed according to Lemma A.1. Let $Z_n$ be the periodic extension of $R_n$. Then $\rho(Z_n)=\rho_n$, and by part (ii) of Lemma~\ref{lemma-periodic-GSC},
\be\label{epn}
\lim_{n\to\infty}\frac{e_p(X_n)}{\rho_n}\leq\lim_{n\to\infty}\frac{e_p(Z_n)}{\rho_n}={v(\0)\over 2}-{|v(\q)|\over 4}.
\ee
Applying Eq.~(\ref{epix-eps}) to $X_n$,
\be
\frac{e_p(X_n)}{\rho_n}+\frac{u(\0)}{2\rho_n}=\omega_{\Lambda_n}(X_n\cap\Lambda_n) =\frac{1}{2V(\Lambda_n)}\int_{\Lambda_n}\int_{\Rr^d}u(\x-\y)\d\mu_{X_n}(\y)\d\mu_{X_n}(\x),
\ee
where $\Lambda_n$ is any period parallelepiped of $X_n$. Suppose that $\mu_{X_n}$ tends to $\lambda$ on Schwartz functions. If $u\in\mathcal{S}(\Rr^d)$, this would yield $v(\0)/2$ instead of (\ref{epn}). If $u$ is not a Schwartz function, replace it by $u*G^t$, compute the limit of $e_p(X_n)/\rho_n$ and let $t$ tend to zero. However,
\be
\int_{\Rr^d}(u*G^t)(\x-\y)\d\y=v(\0)
\ee
for any $t\geq 0$; thus, again, we obtain a contradiction. $\Box$

\newsec{Interactions without Bravais lattice ground states at high density}\label{sec-noBravais}

For certain potentials having a partially negative Fourier transform we can complete the information about GSC's given in part (ii) of Theorem~\ref{hom-inhom}. Namely, we will show that for these potentials no periodic GSC $X$ can be a singly or multiply occupied Bravais lattice if $\rho(X)$ is large enough, that is, $J>1$ must hold and all $\y_j$ cannot coincide [cf. (\ref{Xgen})]. This result is valid in any dimension. Our method is to exclude Bravais lattices first as IDGS's and then as GSC's on the torus and in infinite space at high but finite densities. We note that one-dimensional examples of interactions having a nondegenerate GSC with two points in the unit cell~\cite{Ven,NiR} and others with no periodic GSC~\cite{HaR} existed already thirty years ago, see also Radin's review~\cite{Rad}.

\subsection{Compensation by higher harmonics}

In the following example the negative contribution to the energy coming from a wave vector $\k$ is compensated by the positive contribution coming from integer multiples of $\k$.

\begin{theorem}\label{theor-compensation}
Consider a bounded integrable pair potential having a partly negative Fourier transform $v$ with the property that for any $\k\neq 0$
\be
\sum_{n=1}^\infty v(n\k)\geq 0.
\ee
Then no Bravais lattice can be an IDGS: if $B$ is a Bravais lattice and the parallelepiped $\Lambda$ is a period cell for $B$, then $I[\mu_{B\cap\Lambda}]>I^0$. Furthermore, at finite, high enough densities there is no Bravais lattice among the ground state configurations on tori and in infinite space, and no lattice tower of the form
\be\label{X/B}
B^J=\biguplus_{j=1}^J B
\ee
where $B$ is a Bravais lattice and $J\geq 1$ can be a GSC if $\rho=J\rho(B)$ is high enough. How large $\rho$ must be depends on the details of the interaction. As an example, suppose that\\
(i) there exist $0<k_1<k_2\leq 2k_1$ such that $v(\k)<0$ for $k_1<|\k|<k_2$ and $v(\k)\geq 0$ otherwise,\\
(ii) $v(2\k)\geq |v(\k)|$ for $k_1<|\k|<k_2$, and\\
(iii) $v(\k)=0$ for $|\k|\geq 3q$ with some $q$ such that $k_1<q<k_2$.\\
Then there is no Bravais lattice GSC if
$$\rho\geq (3/2)^{d-1}(q/\pi)^d.$$
\end{theorem}
\emph{Proof.} Consider Equation~(\ref{I[muB]}). If $\K\in B^*$ then $n\K\in B^*$ as well. Summing over $B^*$ along lattice half-lines, the contribution of each partial sum is nonnegative, which shows that $I[\mu_{B\cap\Lambda}]\geq v(\0)/2$. On the other hand, choosing any $\q$ such that $v(\q)<0$, $I^0\leq I[\mu_\q]=v(\0)/2-|v(\q)|/4$, cf. Eq.~(\ref{Imuq}), thus, $\mu_{B\cap\Lambda}$ is not an IDGS. The absence of Bravais lattice ground states at high but finite densities is true both in finite and infinite volume. To see this, suppose that $\Lambda$ is a period cell of $B$. Then
\be
I[\mu_{B^J\cap\Lambda}]=\omega(B^J\cap\Lambda)=\omega(B\cap\Lambda)\geq \frac{v(\0)}{2}.
\ee
On the other hand, if $\rho$ is high enough then according to Lemma A.1 and due to the continuity of $I[\mu]$ one can choose a configuration $R=\left(\r_1,\ldots,\r_{|B^J\cap\Lambda|}\right)\subset\Lambda$ such that
$$|I[\mu_{R}]-I[\mu_\q]|<|v(\q)|/8$$
and therefore
\be\label{RXLambda}
I[\mu_R]=\omega(R)<\frac{v(\0)}{2}-\frac{|v(\q)|}{8}\leq \omega(B^J\cap\Lambda)-|v(\q)|/8.
\ee
This shows that $B^J\cap\Lambda$ cannot be a ground state on $\Lambda$. In infinite space we compare the energies per particle $e_p(B^J)$ and $e_p(R_{\rm per})$ where $R_{\rm per}$ is the periodic extension of $R$. Now $B^J$ and $R_{\rm per}$ have the same density, and Eqs.~(\ref{epix-eps}) and (\ref{RXLambda}) imply $e_p(R_{\rm per})<e_p(B^J)$. From Lemma~\ref{lemma-periodic-GSC} we conclude that $B^J$ is not a GSC.

Consider the specific example. The idea is to find a periodic discrete point approximation $X$ of the density $1+\cos\q\cdot\x$ such that $\omega(X\cap\Lambda)<v(\0)/2$. The computation is done by using the expression
\be\label{epsFour}
\omega(\r)_N=\frac{1}{2}\sum_{\k\in\Lambda^*}v(\k)\left|\frac{1}{N}\sum_{n=1}^N e^{i\k\cdot\r_n}\right|^2,
\ee
see Eqs.~(\ref{omegarn}) and (\ref{exp}). Select a $\q$ whose length $q$ satisfies condition~(iii). Choose $L$ so that $q=(2\pi/L)l$ where $l$ is an integer. Let $\e_j$, $j=1,\ldots,d$ be the Cartesian unit vectors chosen in such a way that $\q=q\e_1$, and
\be
\Lambda=\left\{\sum_{j=1}^d x_j\e_j: 0\leq x_j<L\right\}.
\ee
For a positive integer $l_\perp$, with $a=L/l_\perp$ define the lattice
\be
B_q=\left\{\frac{2\pi}{q}n_1\e_1+a\sum_{j=2}^d n_j\e_j : (n_1,\ldots,n_d)\in\Zz^d\right\}
\ee
and, for any integer $M\geq 2$, the periodic configuration
\be
X=\bigcup_{m=1}^M(B_q+\y_m)
\ee
where
\be
\y_m=\frac{m-1}{M}\frac{\pi}{q}\,\e_1,\quad m=1,\dots,M.
\ee
With this choice of $L$ and $a$, the reciprocal lattice
\be
B_q^*=\left\{n_1\q+\frac{2\pi}{a}\sum_{j=2}^d n_j\e_j : (n_1,\ldots,n_d)\in\Zz^d\right\}
\ee
is a subset of $\Lambda^*=(2\pi/L)\Zz^d$. We show that for $\rho\geq (3/2)^{d-1}(q/\pi)^d$ the energy per particle of $X$ is lower than that of any $B^J$ of the same density. Now
\bea
\frac{1}{|X\cap\Lambda|}\sum_{\x\in X\cap\Lambda}e^{i\k\cdot\x}&=&\frac{1}{|B_q\cap\Lambda|}\sum_{\r\in B_q\cap\Lambda}e^{i\k\cdot\r}\frac{1}{M}\sum_{m=1}^M e^{i\k\cdot\y_m}\nonumber\\
&=&\sum_{\K\in B_q^*}\delta_{\k,\K}\frac{1}{M}\sum_{m=1}^M e^{i\k\cdot\y_m}
\eea
if $\k\in\Lambda^*$, therefore the sum (\ref{epsFour}) reduces to $B_q^*$. The contribution of $\k=\0$ to $\omega$ is $v(\0)/2$. For $\k=\pm\q$ we have
\be\label{sum}
\left|\frac{1}{M}\sum_{m=1}^M e^{i\k\cdot\y_m}\right|^2=\frac{2}{M^2\left(1-\cos\frac{\pi}{M}\right)}
\left\{\begin{array}{ll}
=\frac{1}{2},&M=2\\
\to\left(\frac{2}{\pi}\right)^2,& M\to\infty
\end{array}\right.
\ee
and the function decreases monotonically with $M$ to its limit at $M=\infty$. For $\k=\pm 2\q$, $\exp(i\k\cdot\y_m)$ runs over the $M$th roots of unity and, therefore,
$
\sum_{m=1}^M e^{i\k\cdot\y_m}=0.
$
Under the condition that all other vectors of $B_q^*$ have a length not smaller than $3q$,
\be\label{periodic}
\omega(X\cap\Lambda)=\frac{v(\0)}{2} - \frac{2|v(\q)|}{M^2\left(1-\cos\frac{\pi}{M}\right)}.
\ee
The density of $X$ is
\be
\rho=\frac{M q}{2\pi a^{d-1}}.
\ee
In one dimension $B_q^*=q\Zz$, so the condition of the validity of (\ref{periodic}) is satisfied, and the smallest density to which (\ref{periodic}) applies is $q/\pi$, obtained with $M=2$. The density can be increased only by increasing $M$, therefore the pair energy (\ref{periodic}) increases with the density, but for all $\rho\geq q/\pi$
\be
\omega(X\cap\Lambda)\leq\frac{v(\0)}{2}-\left(\frac{2}{\pi}\right)^2|v(\q)|,
\ee
where the upper bound is obtained by substituting (\ref{sum}) with its limit at $M=\infty$. For $d\geq 2$ we can reach the smallest value for both $\omega(X\cap\Lambda)$ and $\rho$ if we choose the smallest $M$, that is, $M=2$, and the largest $a$ compatible with the condition of validity of (\ref{periodic}). For $a$ the condition implies $2\pi/a\geq 3q$, so its largest allowed value is $a= 2\pi/3q$. To summarize, with (\ref{epix-eps}) we find that the energy per particle of any infinite-volume ground state configuration $G$ of density $\rho\geq (3/2)^{d-1}(q/\pi)^d$ satisfies the inequality
\be
e_p(G)\leq\frac{\rho v(\0)}{2}-\frac{u(\0)}{2}-\frac{\rho |v(\q)|}{2}\leq e_p(B^J)-\frac{\rho |v(\q)|}{2}\quad (d\geq 2)
\ee
and
\be
e_p(G)\leq\frac{\rho v(\0)}{2}-\frac{u(\0)}{2}-\frac{4\rho |v(\q)|}{\pi^2} \leq e_p(B^J)-\frac{4\rho |v(\q)|}{\pi^2}
\quad (d=1)
\ee
if $B$ is any Bravais lattice and $J\geq 1$ such that $J\rho(B)=\rho$. $\Box$

\subsection{Potentials with a cusp at zero}\label{subsec-cusp}

If the integral $\int v(\k)|\k|^2\d\k$ is absolutely convergent, then $u$ is twice continuously differentiable at $\0$, and
\be\label{cusp-0}
-\Delta u (\0) = (2\pi)^{-d}\int v(\k)|\k|^2\d\k.
\ee
If
\be
\lim_{K\to\infty}\int_{|\k|<K} v(\k)|\k|^2\d\k=\infty,
\ee
then $u$ has a cusp at zero. We consider as a special case of a cusp if $u(\r)\to \infty$ as $r\to 0$. The condition (\ref{cusp}) below implies that $u$ has a cusp at zero.

\begin{theorem}\label{theor-cusp}
Let $v^-$ be nonzero and of compact support. Suppose that there exist $K,c>0$ such that
\be\label{cusp}
v(\k)\geq\frac{c}{|\k|^{d+2}}\quad\mbox{for $|\k|\geq K$}.
\ee
Let $\Lambda$ be any parallelepiped. Then, no Bravais lattice commensurate with $\Lambda$ can be an IDGS in $\Lambda$. In particular:

\noindent (i) If $d>1$ and
\be\label{cusp1}
\frac{c}{|\k|^{d+2}}\leq v(\k)\leq\frac{c'}{|\k|^{n+2+\varepsilon}}\quad\mbox{at $|\k|\geq K$}
\ee
for some positive integer $n<d$, positive $c'$ and $0<\varepsilon<1$, then IDGS's may be singular continuous measures of the form
\be\label{mu0}
\mu_0=\lambda_{d-n}\times \mu_{B^{(n)}\cap\Lambda^{(n)}}\ ,
\ee
cf. Eq.~(\ref{lattice}), in which case high-density ground state configurations in $\Lambda$  are strongly anisotropic, and may or may not be Bravais lattices.

\noindent (ii) If $d\geq 1$ and
\be\label{cusp2}
v(\k)\geq\frac{c}{|\k|^3}\quad\mbox{for $|\k|\geq K$},
\ee
then no IDGS is a stationary point of the energy functional. Furthermore, high-density ground state configurations are different from Bravais lattices and their towers (\ref{X/B}) in $\Lambda$ and, if $u$ is bounded, also in infinite space.
\end{theorem}
\emph{Remark.} At the end of Section~\ref{sec-stat} we discussed the case when $u(\x)\sim |\x|^{-d+\zeta}$ with $0<\zeta<d$ near the origin. By a Tauberian argument $v(\k)\sim |\k|^{-\zeta}$ at infinity. If also $v^-\neq 0$, then $d\geq 4$ and $3<\zeta<d$ is covered by (\ref{cusp1}), and $0<\zeta<\min\{d,3\}$ is covered by (\ref{cusp2}). If $0<\zeta\leq 1$ then $\sum_{n=1}^\infty v(n\k)=\infty$ for all nonzero $\k$, which is a special case also of Theorem~\ref{theor-compensation}.

\noindent
\emph{One-dimensional example to (ii).} Let
\be
v(k)=\frac{1}{k^2}e^{-a/k^2}+(1-k^2)e^{-k^2}.
\ee
For $a\geq 3$ this function takes on negative values on two symmetric bounded intervals and decays as $1/k^2$ at infinity. Thus, $u$ is bounded but has a cusp at zero, and
$$\sum_{k\in\Lambda^*}|\widehat{\mu}(k)|^2<\infty$$
must hold for an IDGS. Hence, any IDGS is absolutely continuous in $\Lambda=[-L/2,L/2]$ and is different from the Lebesgue measure, because $v^-\neq 0$. High-density GSC's form a non-arithmetic sequence.

\noindent
\emph{Proof.} Let us return to the tempered measure $G^t_\Lambda*\mu$ and Equation~(\ref{IGt}). For any $t>0$,
\bea
\frac{\d}{\d t}I[G^t_\Lambda*\mu]&=&-\sum_{\k\in\Lambda^*}v(\k)|\k|^2 e^{-2t|\k|^2}|\widehat{\mu}(\k)|^2\nonumber\\
&=&\sum_{\k\in\Lambda^*}v^-(\k)|\k|^2 e^{-2t|\k|^2}|\widehat{\mu}(\k)|^2
- \sum_{\k\in\Lambda^*}v^+(\k)|\k|^2 e^{-2t|\k|^2}|\widehat{\mu}(\k)|^2,
\eea
because both sums are absolutely convergent. Now $\lim_{t\to 0}\d I[G^t_\Lambda*\mu]/\d t$ exists, it can be $-\infty$ for a general $\mu$, but it must be finite nonnegative if $\mu$ is an IDGS. Thus, for any IDGS $\mu$ in $\calm$,
\be\label{condIDGS2}
\sum_{\k\in\Lambda^*}v^+(\k)|\k|^2 |\widehat{\mu}(\k)|^2\leq \sum_{\k\in\Lambda^*}v^-(\k)|\k|^2 |\widehat{\mu}(\k)|^2\leq\sum_{\k\in\Lambda^*}v^-(\k)|\k|^2 <\infty.
\ee
Let $B$ be a Bravais lattice commensurate with $\Lambda$ and let $\mu_{B\cap\Lambda}$ be the associated measure, cf. Eq.~(\ref{muB}). Then
\be
\sum_{\k\in\Lambda^*}v^+(\k)|\k|^2 |\widehat{\mu_{B\cap\Lambda}}|(\k)|^2= \sum_{\k\in B^*}v^+(\k)|\k|^2=\infty
\ee
because of (\ref{cusp}), so $\mu_{B\cap\Lambda}$ is not an IDGS.

Combining (\ref{condIDGS2}) with (\ref{cusp}), we find that
\be
\sum_{\k\in\Lambda^*}\frac{|\widehat{\mu}(\k)|^2}{|\k|^d}<\infty
\ee
must hold for any IDGS $\mu$. Thus, either $\widehat{\mu}$ is nonvanishing on a lower than $d$-dimensional subset of $\Lambda^*$ or it decays to zero.

\noindent
(i) In this case
\be
\sum_{\k\in\Lambda^*:k_1=\ldots=k_{d-n}=0}v^+(\k)|\k|^2<\infty,
\ee
therefore a measure of the form (\ref{mu0})
[which is a stationary point of $I$, c.f. Eq.~(\ref{lattice})] cannot be excluded from being an IDGS. Indeed,
\bea
\sum_{\k\in\Lambda^*}v^+(\k)|\k|^2 |\widehat{\mu_0}(\k)|^2=
\sum_{\k'\in(B^{(n)})^*}v^+(\0_{d-n},\k')|\k'|^2<\infty.
\eea
Now $\mu_0$ can be obtained as the weak limit of a sequence of measures associated with Bravais lattices: the component $\mu_{B^{(n)}}$ is unchanged in this limit, while $\lambda_{d-n}$ is approximated by a sequence of discrete measures associated with Bravais lattices that fill in $\Lambda^{(d-n)}$ more and more densely. The approximating measures can, thus, be associated with more and more anisotropic Bravais lattices.

\noindent
(ii) Combining (\ref{condIDGS2}) with (\ref{cusp2}) one finds
\be\label{muhat-decaying}
\sum_{\k\in\Lambda^*}\frac{|\widehat{\mu}(\k)|^2}{|\k|}<\infty
\ee
for any IDGS $\mu$. Apart from the Lebesgue measure, no stationary point of the energy functional satisfies Eq.~(\ref{muhat-decaying}). Moreover, in contrast with $\lambda$ and $\mu_0$, a $\mu\neq\lambda$ satisfying (\ref{muhat-decaying}) cannot be obtained as the weak limit of measures associated with Bravais lattices. Indeed, for any Bravais lattice $B$, $\widehat{\mu_{B\cap\Lambda}}(\k)=1$ on a sublattice of $\Lambda^*$; therefore, if $\mu\neq\lambda$ was a weak limit of measures associated with Bravais lattices, then $\widehat{\mu}(\k)=1$ would be on lattice, and (\ref{muhat-decaying}) would fail. It follows that there exists some $\rho_\Lambda$ such that if $B$ is a Bravais lattice, $\Lambda$ is a period cell for $B$ and $\rho(B^J)=J|B\cap\Lambda|/V\geq \rho_\Lambda$, then $B^J\cap\Lambda$ is not a ground state of $u_\Lambda$. In one and two dimensions a pair potential can be bounded and satisfy condition (\ref{cusp2}). Then, due to the uniform convergence proven in Corollary~\ref{coro-uniform}, $\rho_\Lambda$ can be chosen independent of $\Lambda$, and we can conclude that in infinite space $B^J$ is not a GSC of $u$. $\Box$

\newsec{The penetrable sphere model}\label{sec-penetrable}

So far we have only given one example, Proposition~\ref{prop-vgeq-piling}, illustrating Corollary~\ref{propCinf} or Proposition~\ref{prop-piling}. However, in that case GSC's of the form of lattice towers | particles superimposed on lattice sites | are submerged in a continuum of other GSC's. The truly interesting situation is when a lattice tower ground state occurs in a non-degenerate manner. Intuition to find such interactions may be based on both $\r$- and $\k$-space considerations. The $\k$-space representation (\ref{nonneg}) of $I[\mu]$ suggests that IDGS's are to be looked for among Dirac combs concentrated on Bravais lattices if $v(\k)\leq 0$ for $|\k|$ large enough. Choosing the density of the lattice $B$ large enough, apart from $\widehat{\mu_B}(\0)=1$, $\widehat{\mu_B}(\k)$ will be nonzero only where $v(\k)\leq 0$, with its modulus equal to 1 on the lattice $B^*$. Thus, one may expect that $I[\mu]$ is  minimized by some high-density $\mu_B$, see also Eq.~(\ref{stability-cond-v}). If $v$ has a unique minimum at $k_m$, the simplest guess is that $B^*$ is close-packed with a nearest-neighbor distance (close to) $k_m$. Thinking in $\r$-space, the best candidates are potentials that are flat or even nesting at zero, that is, having $\Delta u(\0)\geq 0$. The interactions $u(\x)\sim \exp(-\alpha|\x|^\beta)$ with $\beta>2$, studied by Likos \emph{et al.}~\cite{Lik}, belong to this class. Intuitively, such an interaction may prefer the formation of lattice towers. The large negative part of $v$ in such a case, seen on (\ref{cusp-0}), supports this intuition. Below we discuss the penetrable sphere model which is the simplest although somewhat pathological example. A one-dimensional family of pair potentials giving rise to superimposed particles in ground state configurations is presented in a separate publication~\cite{Su4}.

The particles in the penetrable sphere model interact via the pair potential
\be
u(\x)=\left\{\begin{array}{cll}u_0>0&{\rm for}&|\x|<d_0\\
                              0  &{\rm for}&|\x|\geq d_0.
            \end{array}\right.
\ee
We define $u$ to be lower semicontinuous. This potential is applied in soft matter physics to model a system of interpenetrating micelles in a solvent \cite{MW}-\cite{FLL}. The interaction has a partly negative Fourier transform. In 3 dimensions it reads ($k=|\k|$)
\be
v(\k)=\frac{4\pi u_0d_0^3}{(kd_0)^2}\left(\frac{\sin kd_0}{kd_0}-\cos kd_0\right).
\ee
Note that $v$ is not absolutely integrable. The first and deepest minimum of $v(\k)$ is at $k=k_m=6.12/d_0$. If this was to determine the periodicity of the ground state then the ground state would be the dual of an fcc lattice of lattice constant $k_m$, that is, a bcc lattice of nearest-neighbor distance $\sqrt{6}\pi/k_m=(7.70/6.12)d_0$, cf. Eq.~(15) of Ref.~\cite{Su2}. Instead, we will see that the ground state is an fcc lattice of nearest-neighbor distance $d_0$.

We show that with a proper choice of $\Lambda$, the ground state configurations can be given for every particle number $N$. In $d$ dimensions choose $\a_1,\ldots,\a_d$ such that $|\a_j|=d_0$ and the vectors $\a_j$ generate a $d$-dimensional close-packed lattice,
\be
B=\left\{\sum_1^d n_j\a_j: n_1,\ldots,n_d\in\Zz\right\}.
\ee
Choose $\Lambda$ to be a period cell for $B$ containing $|B\cap\Lambda|=M$ points of $B$ and having a side length $\geq 2d_0$ in every direction. The volume of $\Lambda$ is
\be
V=M\left|\det[\a_1\ldots \a_d]\right|,
\ee
so the volume of the unit cell of $B$ is
\be
\rho_B^{-1}=V/M=\left|\det[\a_1\ldots \a_d]\right|=\left\{\begin{array}{cl}
\sqrt{3}d_0^2/2 & \mbox{if d=2}\\
d_0^3/\sqrt{2} & \mbox{if d=3}.
\end{array}\right.
\ee
The property of $B$ is that for the given density $\rho_B$ it has the largest nearest-neighbor distance or for the given nearest-neighbor distance $d_0$ the largest density among Bravais lattices. We make also the hypothesis (trivial in two and proven in three dimensions) that $B$ realizes the densest packing of hard spheres of diameter $d_0$. Note that with the choice of $\Lambda$ we exclude the occurrence of other close-packed (e.g. hcp) structures which can be ground states in suitable domains or in infinite space.

If $N<M$, the ground state in $\Lambda$ is continuously degenerate, any configuration with all inter-particle distances $\geq d_0$ is a ground state. Below we describe the ground state configurations (GSC) for $N\geq M$, modulo cyclic translations in $\Lambda$.

\begin{proposition} The IDGS is $B\cap\Lambda$,
\be
C_\Lambda[u]=\omega(B\cap\Lambda).
\ee
For $N$ finite, let $n\geq 1$ be an integer and let $nM\leq N\leq (n+1)M$. Then there are ${M\choose N-nM}$ GSC's in $\Lambda$. In each of them $(n+1)M-N$ points of $B\cap\Lambda$ are occupied by $n$ particles and $N-nM$ points of $B\cap\Lambda$ are occupied by  $n+1$ particles. Moreover,
\be
\epsilon_N=u_0\frac{nV}{N-1}\left[1-\frac{(n+1)M}{2N}\right],
\ee
so that
\be
C_\Lambda[u]=\lim_{N\to\infty}\epsilon_N=\frac{u_0}{2\rho_B}=\frac{u_0}{2}\times\left\{\begin{array}{cl}
\sqrt{3}d_0^2/2 & \mbox{if d=2}\\
d_0^3/\sqrt{2} & \mbox{if d=3}.
\end{array}\right.
\ee
\end{proposition}

\noindent
\emph{Proof.} The proof goes by induction over $N$. Because each side of $\Lambda$ has a length $\geq 2d_0$, only a single term of $u_\Lambda$ can be nonvanishing, thus, $u_\Lambda(\r)=u_0$ or 0. If $N=M$, $B\cap\Lambda$ has zero energy, therefore it is a GSC. Any perturbation of $B\cap\Lambda$ creates at least one pair of particles of distance $<d_0$, therefore of energy $u_0$. Thus, $B$ is the unique GSC in $\Lambda$. Suppose we know the result up to $N$, where $nM\leq N<(n+1)M$, and want to prove it for $N+1$. Adding a particle to a ground state of $N$ particles costs the less if it is placed on a point of $B\cap\Lambda$ occupied by $n$ particles. The increase in energy is $nu_0$; at any other place it is at least twice as much. No relaxation of the configuration thus obtained can decrease the energy, so it is a ground state for $N+1$ particles. The ground state energy for $N$ particles is
\be
E_0(N)=(N-n M)\frac{n(n+1)}{2}u_0
+[(n+1)M-N]\frac{n(n-1)}{2}u_0
=nu_0[N-\frac{n+1}{2}M].
\ee
Dividing by $N(N-1)/V$ yields $\epsilon_N$. $\Box$

An unpleasant feature of the penetrable sphere model is the discontinuity of its energy at zero temperature when $N\geq M$. The values of $U_\Lambda(\r)_N$ can only be integer multiples of $u_0$ taken from the set $\{E_0(N),\ldots,N(N-1)u_0/2\}$. Therefore,
\bea\label{Zsum}
e^{-\beta [F_N(\beta)-E_0(N)]}&\equiv&\frac{1}{V^N}\int_{\Lambda^N}e^{-\beta [U_\Lambda(\r)_N-E_0(N)]}\d(\r)_N\nonumber\\
&=&\sum_{m=m_0}^{N(N-1)/2}e^{-\beta mu_0}\frac{\lambda_{dN}(A_m)}{V^N}.
\eea
Here $\lambda_{dN}$ denotes the $dN$-dimensional Lebesgue measure,
\be
A_m=\left\{(\r)_N\in\Lambda^N: U_\Lambda(\r)_N=E_0(N)+mu_0\right\},
\ee
and
\be\label{m0}
m_0=\min\{m:\lambda_{dN}(A_m)>0\}.
\ee
Now $A_0$ is the set of translates of $B\cap\Lambda$ increased by the finite number of possible assignments of $N$ particles to $M$ sites, so that $\lambda_d(A_0)>0$, but its higher than $d$-dimensional Lebesgue measures vanish. On the other hand,
\be
\lambda_{dN}(A_{N(N-1)/2})=V\lambda_d(B_{d_0/2})^{N-1}>0,
\ee
where $B_{d_0/2}=\{\x\in\Rr^d:|\x|<d_0/2\}$. Therefore $0<m_0\leq N(N-1)/2$. From (\ref{Zsum}) it follows that
\be
e^{-\beta m_0u_0}\frac{\lambda_{dN}(A_{m_0})}{V^N}\leq e^{-\beta [F_N(\beta)-E_0(N)]}\leq e^{-\beta m_0u_0}.
\ee
Taking the logarithm and dividing by $-\beta $,
\be\label{up-low}
m_0 u_0\leq F_N(\beta)-E_0(N)\leq m_0 u_0+\frac{1}{\beta}\ln\left[V^N/\lambda_{dN}(A_{m_0})\right].
\ee
Letting $\beta$ tend to infinity in (\ref{up-low}) and in the mean energy
\be\label{enT}
E_N(\beta)=-\frac{\partial\ln Z_{\Lambda,N}}{\partial\beta} = E_0(N)+ \frac{\sum_{m=m_0}^{N(N-1)/2}mu_0e^{-\beta mu_0}\lambda_{dN}(A_m)}{\sum_{m=m_0}^{N(N-1)/2}e^{-\beta mu_0}\lambda_{dN}(A_m)},
\ee
we find the following.

\begin{proposition} For the penetrable sphere model in a finite volume,
\be
\lim_{\beta\to\infty}F_N(\beta)=\lim_{\beta\to\infty}E_N(\beta)=E_0(N)+m_0 u_0,
\ee
where $m_0$ is given by Eq.~(\ref{m0}).
\end{proposition}
From the point of view of thermodynamics, it is, therefore, more adequate to consider $E_0(N)+m_0 u_0$ as the ground state energy. The limit of $f_N(\beta)$ as $N$ goes to infinity depends on the limit of $m_0/N^2$ and on the large $N$ behavior of $\lambda_{dN}(A_{m_0})$, and remains to be answered.

\newsec*{Note added}
After the submission of this paper I learned that a part of the tools I developed for studying ground state configurations at high densities existed already in abstract potential theory. The first result dates from 1923 and is due to Fekete~\cite{F}. In an article about the distribution of the roots of polynomials of integer coefficients, Fekete considered the quantity $d_n(S)$, the maximum of the geometric mean of the $n(n-1)/2$ distances between points in an $n$-point set, chosen from an infinite compact set $S$. He proved that $d_n(S)$ was monotone decreasing as $n$ increased, and called the limit the transfinite diameter of $S$. If we take the pair potential $u(\x-\y)=-\ln|\x-\y|$, then the ground state energy per pair of $n$ particles in $S$ is $-\ln d_n(S)$. Thus, Fekete's result is the proof of the monotone increase of the ground state energy per pair [equation~(\ref{monoton})] and of Proposition~\ref{epsilonN} in this special case. What I called the best superstability constant is minus the logarithm of the transfinite diameter. The first proof of the monotone increase of the ground state energy per pair for a general symmetric kernel $u(\x,\y)$ was given by Choquet~\cite{Ch}, who also proved a theorem corresponding to the present Theorem 4.2. The best superstability constant appears in~\cite{Ch} under the name of Fekete constant. The analogous theorem for a kernel which is bounded on the diagonal (Theorem 4.1 here) can be found in a recent paper by Farkas and Nagy~\cite{FN}. Early results on abstract potential theory are summarized in Fuglede's paper~\cite{Fug}.

\newsec*{Acknowledgements}
I thank Szil\'ard R\'ev\'esz for drawing my attention to related results in abstract potential theory.
This work was supported by OTKA Grants K67980 and K77629.

\newsec*{Appendix}
\renewcommand{\theequation}{A.\arabic{equation}}

In this appendix we collect a few results on the convergence of measures on compact sets. Let $\Lambda\subset\Rr^d$ be compact and let $\calm$ be the set of normalized Borel measures on $\Lambda$.

\noindent
\textbf{Lemma A.1} \emph{Any $\mu\in\calm$ is the weak limit of a sequence $\{\mu_{R_N}\}_{N=1}^\infty$ of discrete measures of the form (\ref{gspoint}).}

\noindent
\emph{Proof.} We may suppose that $\Lambda$ is a cube, otherwise, we take a cube that covers $\Lambda$ and extend $\mu$ with zero value to a measure on the cube (cf. Lemma A.4.) We may also suppose that $\Lambda$ is a unit cube. Given $\mu\in\calm$ (thus, $\mu(\Lambda)=1$), divide $\Lambda$ into a disjoint union of $n^d$ semi-open cubes $Q_j^n$ of side length $1/n$, where $n=\lfloor\log_2(N+1)\rfloor$. In $Q_j^n$ choose $N_j$ points $\{\r_{ji}^n\}_{i=1}^{N_j}$ where $N_j=\lfloor N\mu(Q_j^n)\rfloor$ or $N_j=\lceil N\mu(Q_j^n)\rceil$ in such a way that $\sum_jN_j=N$. Thus,
\be
|\mu(Q_j^n)-N_j/N|\leq 1/N.
\ee
Define
\be\label{point}
\mu_{R_N}=\frac{1}{N}\sum_{j=1}^{n^d}\sum_{i=1}^{N_j}\delta_{\r_{ji}^n}
\ee
where $\delta_\r$ is the Dirac measure at $\r$. For any continuous function $f$, let
\be
\langle f\rangle_{n,j}=\frac{1}{\mu(Q_j^n)}\int_{Q_j^n}f\d\mu.
\ee
Because $\Lambda$ is compact, $f$ is uniformly continuous, so there is some $\delta_n$ which goes to zero as $n$ increases, such that for each $j$
\be
\left|\frac{1}{N_j}\sum_{i=1}^{N_j}f(\r_{ji}^n)-\langle f\rangle_{n,j}\right|\leq\delta_n.
\ee
It follows that
\be
\left|\int f\d\mu_{R_N}-\int f\d\mu\right|\leq \delta_n
+\sum_{j=1}^{n^d}\left|\frac{N_j}{N}-\mu(Q_j^n)\right||\langle f\rangle_{n,j}|.
\ee
The second term is $O(n^d/N)$, so both terms tend to zero as $N$ goes to infinity. Thus, $\mu_{R_N}\rightharpoonup\mu$. $\Box$

Next, we show that given any infinite sequence of normalized measures on a compact set, one can select a subsequence converging weakly to a normalized measure. We start with the compact set being a unit cube $\Lambda$ equipped with periodic boundary conditions ($d$-dimensional torus).

The Fourier transform (\ref{muhat}) of a $\mu\in\calm$ is a function of positive type \cite{RSi} on $\Lambda^*=2\pi\Zz^d$, meaning that for any integer $l$, any $\k_1,\ldots,\k_l$ in $\Lambda^*$ and any complex numbers $z_1,\ldots,z_l$
\be\label{posdef}
\sum_{i,j=1}^l \widehat{\mu}(\k_i-\k_j)\overline{z_i}z_j\geq 0.
\ee
It follows among others that
\be
|\widehat{\mu}(\k)|\leq\widehat{\mu}(\0)=1.
\ee

\noindent
\textbf{Lemma A.2} \emph{From any sequence $\mu_n\in\calm$ one can select a subsequence $\mu_{n_i}$ such that $\widehat{\mu}_{n_i}$ converges pointwise to the Fourier transform $\widehat{\mu}$ of a $\mu\in\calm$.
We say that $\mu_{n_i}$ converges to $\mu$ in Fourier transform.}

\noindent
\emph{Proof.}
Because $\Lambda^*$ is a countable set, we can use the diagonal process \cite{Rud} to choose a subsequence  $\widehat{\mu}_{n_i}$ which converges pointwise to some function $\psi$ on $\Lambda^*$,
\be
\lim_{i\to\infty}\widehat{\mu}_{n_i}(\k)=\psi(\k), \quad\mbox{every $\k\in\Lambda^*$}.
\ee
Fixing any integer $l$, $\k_1,\ldots,\k_l$ in $\Lambda^*$ and complex numbers $z_1,\ldots,z_l$, writing down Eq.~(\ref{posdef}) for $\widehat{\mu}_{n_i}$ and taking the limit $i\to\infty$, we find
\be
\sum_{i,j=1}^l \psi(\k_i-\k_j)\overline{z_i}z_j\geq 0.
\ee
Thus, $\psi$ is a function of positive type, $\psi(\0)=1$, and by Bochner's theorem \cite{RSi}, $\psi=\widehat{\mu}$, the Fourier transform of a probability measure $\mu$ on $\Lambda$. $\Box$

\noindent
\textbf{Lemma A.3} \emph{$\mu_n$ converges to $\mu$ in Fourier transform if and only if it converges weakly to $\mu$.}

\noindent
\emph{Proof.} If $\mu_n\rightharpoonup\mu$ then
\be
\lim\mu_n(f)=\mu(f)\quad {\rm for}\quad f(\x)=\sin\k\cdot\x,\ \cos\k\cdot\x,
\ee
therefore $\lim\widehat{\mu}_{n}(\k)=\widehat{\mu}(\k)$ for all $\k\in\Lambda^*$.

Suppose now that $\mu,\mu_n\in\calm$, $\widehat{\mu}=\lim\widehat{\mu}_{n}$. Consider the trigonometric polynomials,
\be\label{tri}
p(\x)=\sum_{\k\in\Lambda^*}a_\k e^{i\k\cdot\x}
\ee
where $a_\k\neq 0$ for a finite number of $\k$. Any continuous function $f$ on $\Lambda$ is the uniform limit of a sequence $p_m$ of the form (\ref{tri}). Given any $\veps>0$, we fix an $m$ so that $\|f~-~p_m\|_\infty\leq\veps$, and choose $N_\veps$ such that for $n>N_\veps$
\be
|\mu(p_m)-\mu_n(p_m)|\leq\sum_{\k\in\Lambda^*}|a^m_\k||\widehat{\mu}(\k)-\widehat{\mu}_{n}(\k)|\leq\veps.
\ee
Because $|\mu(f-p_m)|\leq\veps$ and $|\mu_n(p_m-f)|\leq\veps$, we obtain
\be
|\mu(f)-\mu_n(f)|\leq 3\veps\quad{\rm if}\quad n>N_\veps,
\ee
proving that $\mu_n$ converges weakly to $\mu$. $\Box$

The extension of the above result to the general case, when the compact domain may not be a cube and the boundary condition may not be periodic, is immediate.

\noindent
\textbf{Lemma A.4} \emph{Let $\Lambda_0\subset\Rr^d$ be a compact Borel set and let ${\cal M}_{\Lambda_0}$ denote the set of normalized Borel measures on $\Lambda_0$. Given any infinite sequence $\mu_n\in{\cal M}_{\Lambda_0}$, one can select a subsequence converging weakly to some $\mu\in{\cal M}_{\Lambda_0}$.}

\noindent
\emph{Proof.} Take a cube $\Lambda\supset\Lambda_0$. Let $\mu'_n\in\calm$ with $\supp\mu'_n\subset\Lambda_0$ be the extension of $\mu_n$ to $\Lambda$. Constructing a normalized limit measure $\mu'\in\calm$ according to Lemma A.2, it is clear that $\supp\mu'\subset\Lambda_0$, therefore $\mu=\mu'|_{\Lambda_0}\in{\cal M}_{\Lambda_0}$. Because $\mu'$ is the weak limit of $\mu'_n$, cf. Lemma A.3, $\mu$ is the weak limit of $\mu_n$. $\Box$

\end{document}